\documentclass[conference]{IEEEtran}

\ifCLASSINFOpdf
\else
\fi
\usepackage{newunicodechar}
\newunicodechar{−}{$-$}

\usepackage{verbatim}
\usepackage{url}
\usepackage{todonotes}
\usepackage{array} 
\usepackage{multirow}
\usepackage{booktabs} 
\usepackage{subcaption}
\usepackage{float}
\usepackage{tabularx}
\usepackage{tcolorbox}
\usepackage{adjustbox}
\tcbuselibrary{skins, breakable}

\newtcolorbox{quotebox2}{
  enhanced,
  colback=gray!5,
  colframe=gray!30,
  arc=3pt,
  boxrule=0.05pt,
  drop shadow,
  fontupper=\itshape,
  width=\columnwidth,   
   top=2pt, bottom=2pt,
   before skip=4pt,
    after skip=4pt,
  center
}

\newcommand{\boldtitle}[1]{\vspace{5px}\noindent\textbf{#1}}

\newcommand{\bolditalictitle}[1]{\vspace{5px}\noindent\textbf{\textit{#1}}}

\usepackage{xcolor}
\usepackage{mdframed}
\usepackage{amsmath}
\usepackage{tcolorbox}
\usepackage{enumitem}
\usepackage{arydshln}

\usepackage{url}

\usepackage{breakurl}
\usepackage[breaklinks]{hyperref}

\begin{document}
%
\title{More than Meets the Eye: Understanding the Effect of Individual Objects on Perceived Visual Privacy}


\author{\IEEEauthorblockN{Anonymous Authors}}
	

%


\author{\IEEEauthorblockN{Mete Harun Akcay\IEEEauthorrefmark{1}\IEEEauthorrefmark{2},
Siddarth Rao\IEEEauthorrefmark{1},
Alexandros Bakas\IEEEauthorrefmark{1},
Buse Atli\IEEEauthorrefmark{3}\IEEEauthorrefmark{1}}
\IEEEauthorblockA{\IEEEauthorrefmark{1}Nokia Bell Labs,
Espoo, Finland \\
sid.rao@nokia-bell-labs.com, alexandros.bakas@nokia-bell-labs.com
}
\IEEEauthorblockA{\IEEEauthorrefmark{2}{Å}bo Academy University,
Turku, Finland, meteharun.ackay@abo.fi
}
\IEEEauthorblockA{\IEEEauthorrefmark{3}Linköping University,
Linköping, Sweden, 
buse.atli@acm.org
}
}



\maketitle

\begin{abstract}
User-generated content, such as photos, comprises the majority of online media content and drives engagement due to the human ability to process visual information quickly. Consequently, many online platforms are designed for sharing visual content, with billions of photos posted daily. However, photos often reveal more than they intended through visible and contextual cues, leading to privacy risks. Previous studies typically treat privacy as a property of the entire image, overlooking individual objects that may carry varying privacy risks and influence how users perceive it. We address this gap with a mixed-methods study (n = 92) to understand how users evaluate the privacy of images containing multiple sensitive objects. Our results reveal mental models and nuanced patterns that uncover how granular details, such as photo-capturing context and co-presence of other objects, affect privacy perceptions. These novel insights could enable personalized, context-aware privacy protection designs on social media and future technologies.
\end{abstract}


%
\IEEEpeerreviewmaketitle


\section{Introduction}

\begin{figure*}[t]
    \centering
    \includegraphics[width=\textwidth]{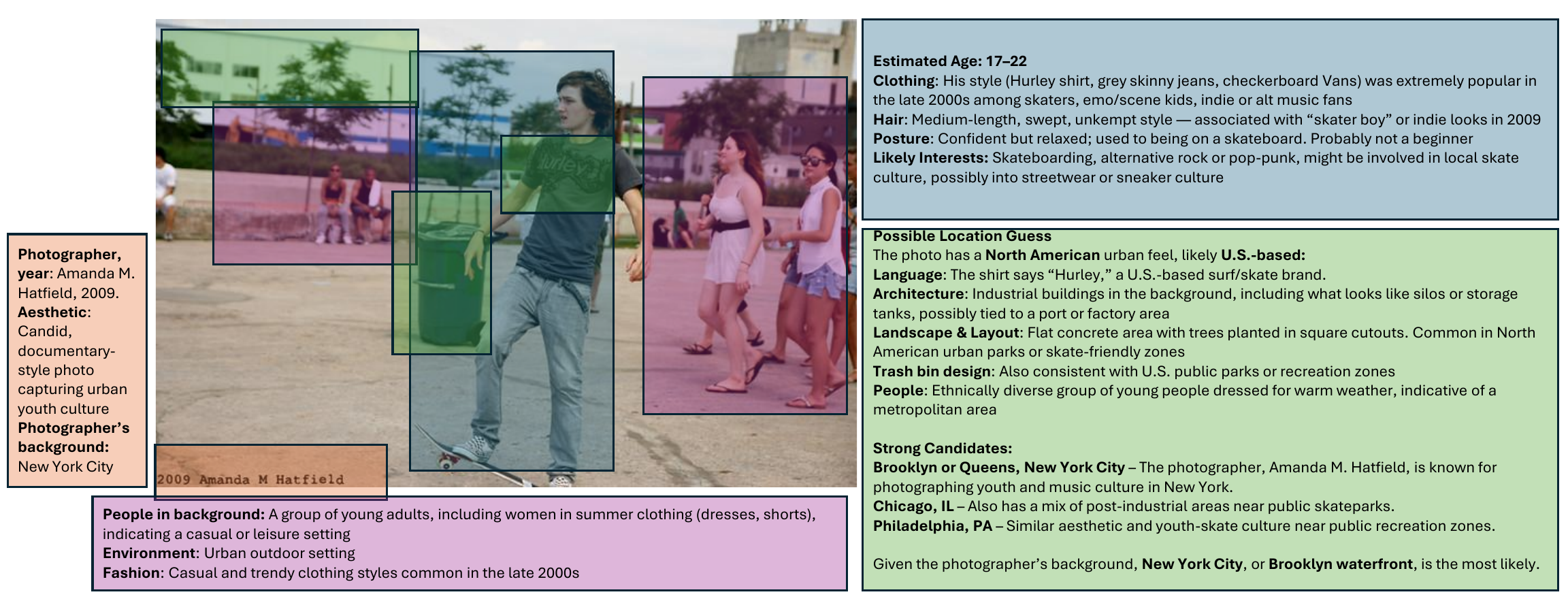}
    \caption{An example demonstrating the amount of information revealed (intentionally or inadvertently) through a single visual content. The image is sampled from the \href{https://cocodataset.org/\#explore}{MS-COCO dataset explorer} 
    and the text descriptions are a combined interpretation by a human observer (one of the authors of this work) and commercial GenAI methods (link to actual image is \href{http://farm3.staticflickr.com/2596/3855197434_72ccaa2ed3_z.jpg}{here}).
    %
    }
    \label{fig:leakage}
\end{figure*}



With the advent and ubiquity of affordable camera technologies, capturing and sharing visual content has become the new norm of everyday social interactions. A study in 2025~\cite{photutorial2023} reports that 14 billion images are shared daily on social media networks and instant messengers, a high percentage of which include photos of people with friends and family members. However, by sharing photos of their private lives and personal moments, they may inadvertently jeopardize their privacy. Exposing personally identifiable or sensitive information online can have serious consequences, such as identity theft~\cite{burnes2020risk},
cyber bullying~\cite{Chan03042019}, or stalking~\cite{kaur2021systematic}.

One of the well-studied privacy concerns related to sharing content online is behavioral profiling and targeted advertisements. These systems mainly rely on users' search queries, browsing patterns, and metadata to transform personal data and privacy into a commodity~\cite{boerman2017online,puglisi2017web}. However, a significant amount of information can be harvested purely from the visual content by humans and computer vision systems. 
Figure~\ref {fig:leakage} demonstrates an example of how much information a single photo can reveal by combining different visual cues extracted from the image. 
%
%
News media investigations~\footnote{\href{https://www.nytimes.com/spotlight/visual-investigations}{The New York Times visual investigations} and \href{https://www.bellingcat.com/category/news/?fwp_categories=news}{Bellingcat investigations}} have demonstrated how open-source intelligence tools and techniques can be utilized to extract rich information from shared visual content that was not intended initially~\cite{ww2photomystery2023,chronolocation2023,ChinaShips2020}.
%
Similarly, accessibility-focused research works in the field of human-computer interaction (HCI) have also explored privacy concerns associated with intentional or accidental disclosure of visual information~\cite{akter2020uncomfortable, akter2020privacy, stangl2020visual, zhang2024designing}. 
However, there is still some uncertainty about whether and how users can understand the complex visual and contextual cues in the photos they share online. Our work aims to explore this research theme to uncover the nuances of perceived visual privacy.

Our goal is to understand how individual objects (i.e., the visual elements within the image), which may have varying privacy risks in different contexts, influence users' privacy decisions. Previous studies focused mainly on the perceived privacy of the entire image to understand users' perceptions and behaviors while sharing photos online. More specifically, these works have focused on understanding whether a user evaluates the whole image as private or not ~\cite{patwari2024Perceptanon, Tonge2020Image,habib2019impact,ahern2007over}, or focuses on assessing the privacy risks of a single object~\cite{orekondy2017towards}. However, these approaches overlook the effect of context and the co-presence of different privacy-sensitive objects within an image. A few works on users' privacy preferences regarding the sharing of visual content on social media have examined spatial context (location where the photo was taken) and social context (e.g., the number of individuals present in the image and the subject's relation to the image)~\cite{hoyle2020privacynorms}. Our work extends this line of research by exploring more fine-grained aspects of visual content.

\boldtitle{Contributions:} 
%
We offer novel insights into how end-users evaluate visual privacy and the factors that shape their privacy perceptions. 
Secondly, we provide empirical evidence on users' privacy heuristics, derived from a mixed-methods analysis of qualitative and quantitative data gathered through an online user survey (n = 92). 
Finally, we show the feasibility of using synthetic images in user studies, as a valuable alternative when real-world data is difficult to obtain. 

\boldtitle{Overview of results:} 
We draw results by fixing the individual visual element (referred to as \textit{objects}) in the foreground and studying its perceived privacy by the user using varying combinations of objects in the background. In summary, we make the following observations.
\begin{itemize}[leftmargin=*]
    \item We found that end-users demonstrate a latent cognitive model for evaluating the intricate visual and contextual cues of the images they publish online. Users pay attention to granular details about the objects, such as what or who is in the photo and the situational context in which the photo is captured. This finding supports the well-established theories of contextual anchors of privacy decisions~\cite{nissenbaum2004privacy}, and extends the knowledge by providing evidence about micro-details of context.
    %
    \item We found that having certain categories of sensitive objects in the background 
    dictates the privacy perception of those in the foreground. Especially when the background objects are unrelated to each other, the presence of other background objects has a negligible impact on the users' privacy perception of the foreground object due to the \textit{dominance effect}. While previous studies have discussed the impact of similar salient features on privacy perception using anecdotal examples, we provide empirical evidence for the dominance effect.
    %
    \item Certain combinations of semantically related sensitive objects in the background contribute to the perceived privacy of those in the foreground. While such individual background objects may still influence users' privacy judgments to some extent, their co-presence can trigger stronger concerns than what each would cause in isolation. %
    Prior research has 
    investigated the inference risks that arise when related elements from different datasets. However, a similar effect emerging from a single data item (such as individual images) is less explored to the best of our knowledge. In this direction, our findings offer novel insights into \textit{pair-wise co-presence effect} and its potential to trigger privacy concerns. 
\end{itemize}

Our work contributes towards a comprehensive understanding of visual privacy perception that can potentially be utilized to develop personalized privacy preference options for sharing visual content online. In particular, the insights from this study can be leveraged for designing assistive technologies that share the entire content while removing, blurring, or masking privacy-sensitive objects, but still function effectively. Such technologies can be integrated into various AI applications, including privacy-sensitive visual aid devices, smart home assistants, wearable cameras, augmented reality platforms, surveillance systems, and live classroom environments. 

\section{Related Work}
\label{sec:Background}

\subsection{Contextual Foundations of Visual Privacy}
Many privacy theories argue that human perception of privacy is both content- and context-dependent~\cite{wisniewski2022privacy}. Rather than treating privacy as a static property of information, these theories highlight how people interpret its meaning, intent, and appropriateness in the moment. For example, contextual integrity theory argues that privacy is about whether information flows align with the social norms of a given context, such as who receives the information, what is being shared and how it is shared~\cite{nissenbaum2004privacy}. Similarly, boundary regulation~\cite{petronio2002boundaries} and privacy calculus~\cite{dinev2006extended, acquisti2007can} treat privacy as a negotiation outcome of interpersonal boundaries with others and of perceived benefits of information disclosure. 
Empirical HCI research supports these theoretical perspectives, showing that users' information sharing decisions depend heavily on situational norms and social relationships. Prior work on visual privacy has explored, e.g., how user dispositions (e.g., demographics, personal traits), the aesthetics of the content (e.g., location, people), the intended audience, the presence of bystanders, and the surrounding environment in which a photo is taken can influence privacy perceptions~\cite{habib2019impact,ahern2007over,hoyle2020privacynorms,Niu2025Bystanders,kairam2016snap}. Collectively, these findings indicate that visual privacy judgments emerge from contextual signals and micro-details of the content.
Our work builds on these foundations and empirical results by examining privacy perception under controlled manipulation of micro-details and contextual conditions. While prior work has focused mainly on whole image, we investigate how object-level cues and their interactions contribute to perceived privacy.

\subsection{Visual Privacy Risks and Protection}
A wide range of privacy risks associated with visual content have been documented in the research literature. Such risks emerge from the distinct and recognizable elements contained in the image, often revealing more information than the sharer intends. Prior work has shown that photos can expose biometric or demographic details, geolocation, social relationships and status, or health and economic conditions~\cite{li2020towards,10.1145/3547299,goyeneche2024linked}. Individuals who did not voluntarily participate in image creation (e.g., bystanders, passersby, or children) could also be exposed when the image is shared without their knowledge or consent~\cite{akter2020uncomfortable,akter2020privacy,Niu2025Bystanders,10.1145/3517384}. Privacy risks are further amplified by sensitive attributes inferred at scale from image content and metadata by modern data-harvesting systems~\cite{amil2024impact}. These findings illustrate that privacy risks in images arise from inferences that can be drawn from how visual and contextual cues relate to each other within a scene.

To mitigate the risks mentioned above, prior work has explored privacy-enhancing technologies for visual content~\cite{10.1145/3708501,wen2024image,chiang2025understanding}. Technical solutions include automatic detection and obfuscation of sensitive details using blurring, redaction, and facial de-identification~\cite{orekondy2017towards,ravi2024review,khamis2022deepfakes,vishwamitra2017blur}. On the other hand, user-centric solutions offer options to restrict the sharing of visual content to a specific set of people or to delegate privacy decisions to trusted assistants~\cite{zhang2024designing,hoyle2020privacynorms, 10.1145/3517384, 10.1145/2396761.2398735}. 
%
However, these solutions treat all sensitive elements as uniformly risky, overlooking how the co-presence of multiple elements impacts users' privacy perception. 
The mismatch between existing protection mechanisms and human perception emphasizes the need for a deeper understanding of object-level interactions, which is the focus of our work.

\subsection{Visual Privacy Methodological Approaches}
Research on visual privacy has traditionally relied on natural images collected from social media, photo-sharing platforms, or curated datasets. Prior studies have used such images in user studies, where participants are asked to report, e.g., their comfort level in sharing them with others, their intended audience, the control mechanisms they use, and the reasoning behind their decisions, which stem from participants' privacy preferences. Furthermore, many studies have leveraged photos provided by participants to investigate which types of content are considered sensitive. Researchers have also studied how the visibility of certain objects affects privacy judgments by obfuscating sensitive regions of the image before measuring participants' privacy perceptions~\cite {orekondy2017towards,li2017effectiveness,khamis2024perspectives}.
Despite ecological validity and high-level insights, the main limitation of using natural images is the lack of control over modulating specific visual aspects. As a result, it is difficult to observe the role of individual foreground or background objects, their co-presence and interactions, or changes in the surrounding environment that shape perceived privacy. 
To overcome such limitations, synthetic images produced by modern generative models that offer the controlled manipulation of visual features can be a helpful approach. Recent HCI research has explored whether humans can differentiate between synthetic and real faces~\cite{lu2023seeing, lago2021more,nightingale2021synthetic,shen2021study}. 
Neurophysiological studies also demonstrate that synthetic images can approximate human perceptual judgments in controlled settings and motivate the use of synthetic stimuli in experimental research~\cite{bilucaglia2024emotional}. Building on this direction, our study employs carefully designed prompts to control the components of images during the image generation process. We use these synthetically generated images as stimuli in a user study aimed at investigating how object-level features and contextual factors shape privacy perception.




\section{Methodology}


\subsection{Preliminaries}
\label{sub:preliminaries}

\newcounter{rq}
\newcommand{\RQ}[2]{\refstepcounter{rq}\textbf{H\label{#1}\therq}~#2}


In line with prior work~\cite{akter2020uncomfortable, Niu2025Bystanders, orekondy2018connecting}, we define the key concepts and their relationships as follows.

\begin{itemize}[leftmargin=*]
     \item \textbf{Object} denotes a bounded, semantically meaningful entity that belongs to a specific category and can be distinguished from other objects and the background. 
     The visual objects (e.g., face, person, tree) are entities that can be recognized based on their visual appearance such as shape, texture, color, while textual objects (e.g., name, date) consist of symbolic representations (characters, words, phrases). 
    \item \textbf{Privacy-sensitive object (PSO)} refers to an object that contains personally identifiable or sensitive information that owners may feel uncomfortable sharing on public platforms. 
    A visual content generally consists of objects with different levels of sensitivity, along with the background. An object may be classified as a PSO either due to explicitly defined data protection regulations or visual indicators such as the surrounding context. As illustrated in Figure~\ref{fig:leakage}, the person in the center of the image and the nearby bystanders are marked as PSO (since they constitute personally identifiable information under GDPR), whereas the sky and the floor belong to background. Other objects, such as trees or trash bins, can also be elevated to PSO when the goal is to infer the geographical location where the image was captured.
    \item \textbf{Foreground PSO} is the primary subject of interest and is typically located in the center of visual content, despite the presence of other sensitive elements in the background. 
    In Figure~\ref{fig:leakage}, the person in the center is the foreground PSO and the bystanders are the background PSO. 
    \item \textbf{Perceived privacy level (PPL)} is the degree to which individuals subjectively believe their personal information, activities, or beliefs are protected from unwanted observation, access, or misuse. 
    \item \textbf{Comfortability level} is the degree of ease and safety experienced by an individual in a given setting. PPL and comfortability level are conceptually aligned, but they are inversely related: as the perceived privacy level increases, the comfortability level decreases. 
\end{itemize}
Based on this terminology, we construct hypotheses as follows:


\noindent \RQ{rq:bg}{: The perceived privacy level of a foreground PSO increases when it is accompanied by a background PSO.}

\noindent \RQ{rq:env}{: The surrounding visual scene affects the perceived privacy level of the same PSO present in the image.}


\noindent  \RQ{rq:co}{: The co-presence of multiple background PSOs further elevates the perceived privacy level of a foreground PSO compared to when only a single background PSO is present.}

\subsection{Study Design}
\label{sec:study_design}

\subsubsection{Selection of Environments}

%
We consider two survey environments: a café and an office. These represent distinct social and professional contexts, both communal spaces where photos may be taken. Privacy expectations are usually lower in cafés due to the sense of privacy through anonymity. In offices, clear roles and the handling of sensitive or private information can make people uncomfortable with taking and sharing photos on social networks.

\subsubsection{Foreground and Background PSOs}
To determine both foreground and background PSOs, we examined the subset of the VISPR dataset designed for image redaction (VISPR-Redacted~\cite{orekondy2018connecting}) and the VizWiz-Priv dataset~\cite{Gurari2019VizWiz}. VISPR-Redacted contains user uploaded, publicly available Flickr images and is curated for automatic detection of PSOs and redacting them by masking. VISPR-Redacted contains 24 labels, all marked as PSOs. 
VizWiz-Priv is a large-scale collection of real-world images captured by blind photographers and accompanied by questions and crowd-sourced answers. VizWiz-Priv includes 23 categories that annotators label as PSO. 
%
Our analysis revealed that both datasets include several identical labels, providing insights into PSOs that are commonly shared with others and online.
We chose the three most common object categories from VizWiz-Priv as our foreground PSOs: \textbf{face, miscellaneous paper and computer screen}. Each foreground PSO carry inherent sensitive information while also offering potential information for users:  
\begin{enumerate}[leftmargin=*]
    \item \textbf{Face}: Visual identification of a person. Sharing a face can support social connection (sending it to friends), identity verification (online applications), or professional representation (LinkedIn profile, news). 
    \item \textbf{Miscellaneous Paper}: Documents such as tickets, forms, printouts, or receipts. These may contain private information (e.g., names, addresses, birthdates, signatures), but are often shared for practical purposes such as proof of purchase, reimbursement, or customer support. 
    \item \textbf{Computer/phone screen}: Digital content displayed on electronic devices. Screens may reveal sensitive information (e.g., usernames, phone numbers, emails), but screenshots are frequently shared for collaboration or troubleshooting purposes.
\end{enumerate}

For each foreground PSO, we identified five background PSOs through an internal expert discussion on both datasets, considering diversity in modality (visual vs.\ textual), potential sensitivity, and their frequency of being together with foreground PSOs. 
Our set of foreground/background PSO combinations is presented in Table~\ref{tab:survey_PSOs}. 

\begin{table}[t]
    \centering
    \caption{List of Foreground and Background PSOs.}\label{tab:survey_PSOs}
    \resizebox{0.9\columnwidth}{!}{
    \begin{tabular}{cc}
        \toprule
        \bf Foreground PSO & \bf Background PSOs\\
        \midrule
        \multirow{2}{*}{Face} & Face (another, a.), Poster, \\
        & Medicine, Tattoo, Landmark \\
        \hdashline
        Miscellaneous  & Face (photo, p.), Full name, \\
        paper & Address, Birth date, Signature \\
        \hdashline
        Computer/phone screen & Face (reflection, r.), Date, \\
        (screen) & E-mail, Username, Phone number \\
        \bottomrule
    \end{tabular}
    }
\end{table}

\subsubsection{Image Generation}
%
%
To prevent participants’ judgments from being influenced by uncontrolled factors such as inconsistent lighting, backgrounds, or photographic context, we deliberately avoided real images. Instead, we relied on synthesized images to keep visual conditions consistent across variations, allowing participants to focus on the intended differences between foreground and background PSOs.

\begin{figure*}[h]
    \centering
    \begin{subfigure}[t]{0.27\textwidth}
        \centering
        \includegraphics[width=\textwidth]{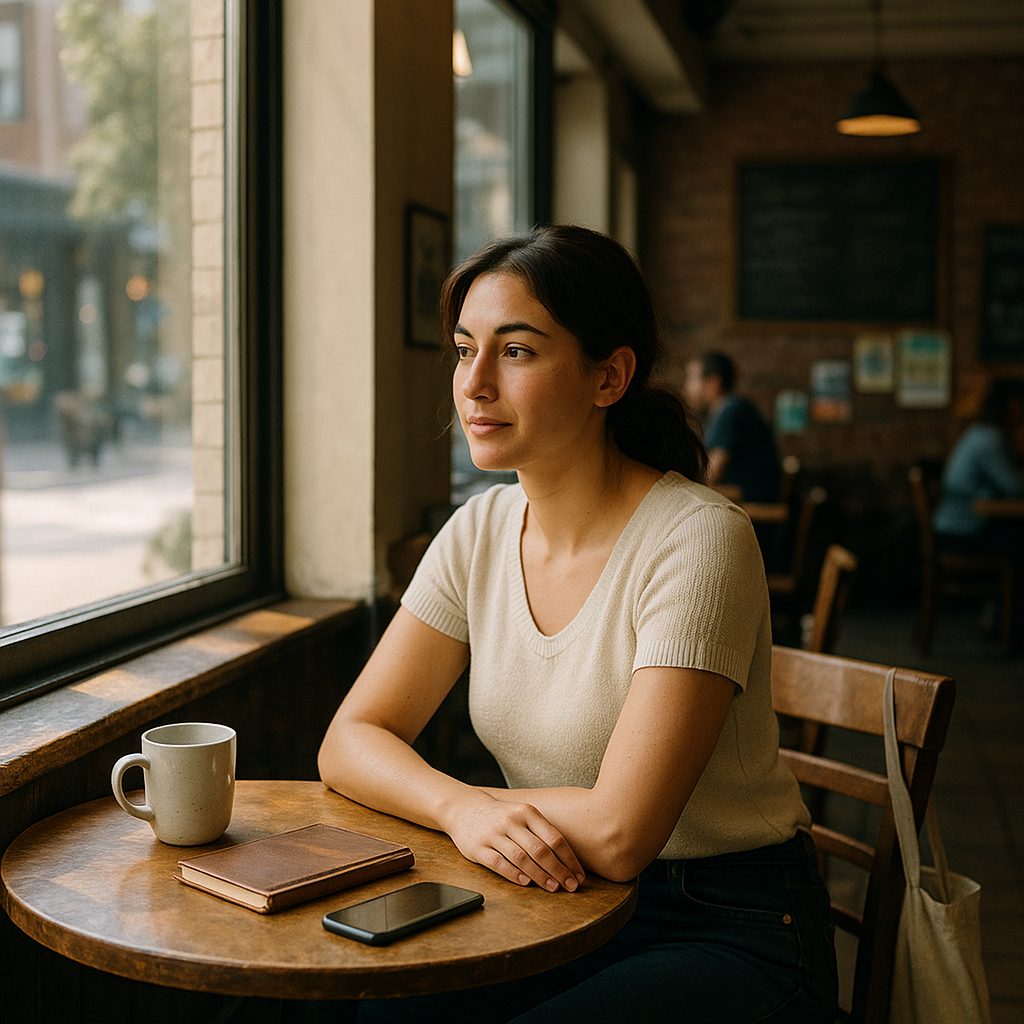}
        \caption{Café: Face only (base image).}
    \end{subfigure}
    \begin{subfigure}[t]{0.27\textwidth}
        \centering
        \includegraphics[width=\textwidth]{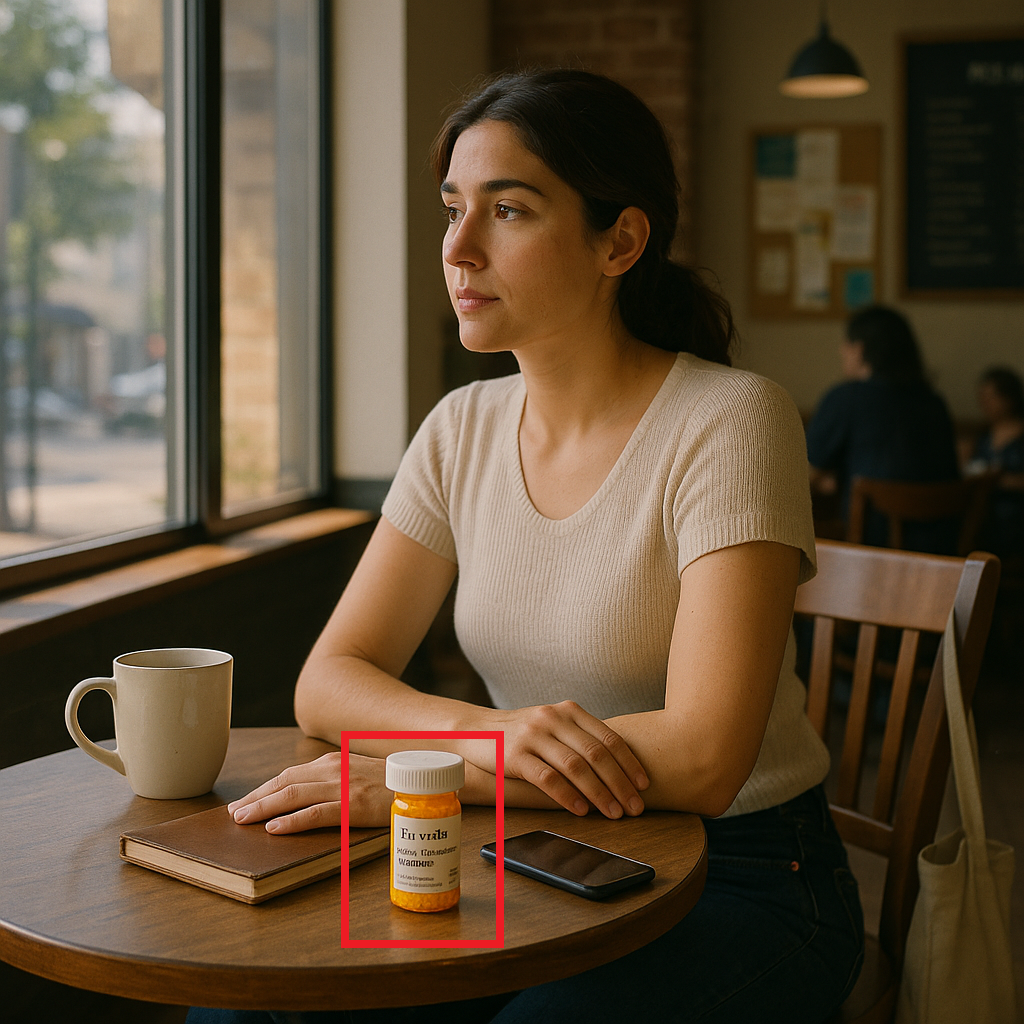}
        \caption{Face + Medicine.}
    \end{subfigure}
    \begin{subfigure}[t]{0.27\textwidth}
        \centering
        \includegraphics[width=\textwidth]{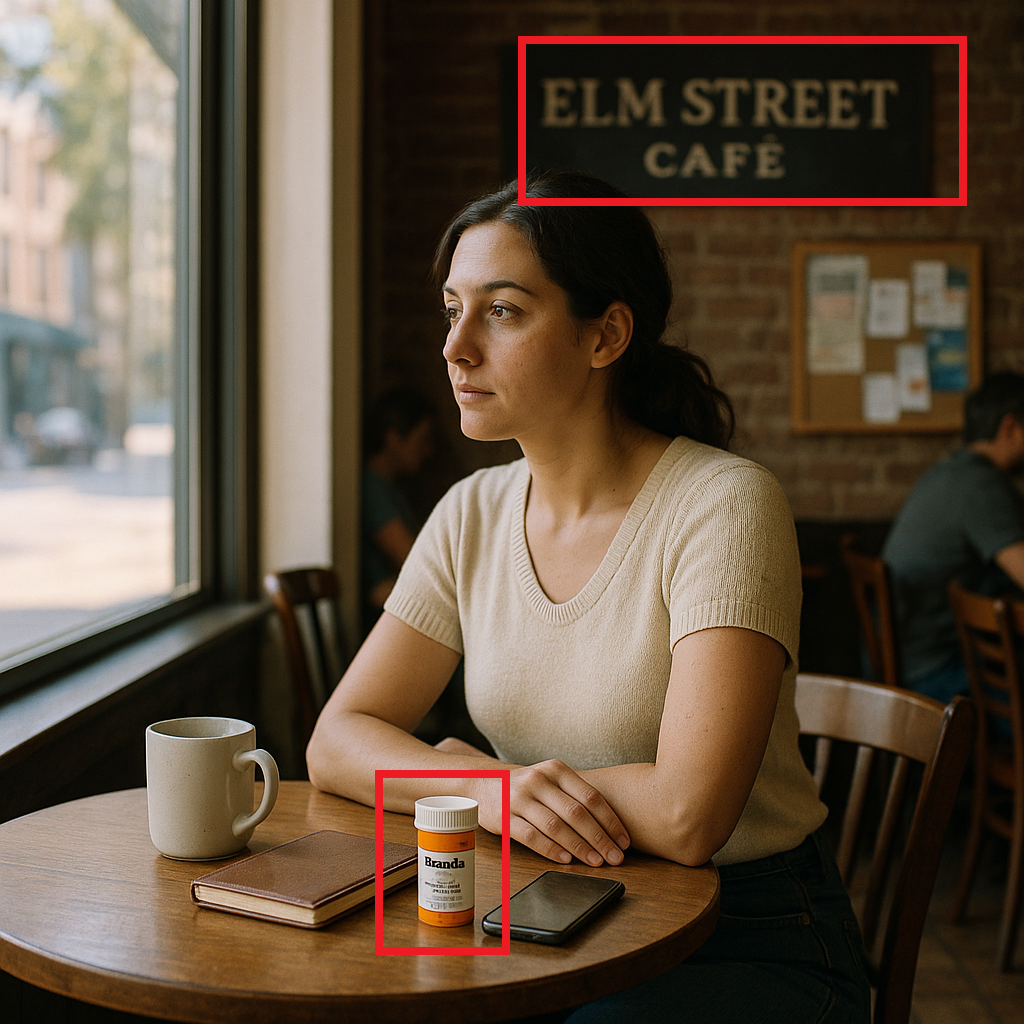}
        \caption{Face + Medicine + Landmark.}
    \end{subfigure}

    \vspace{1em}

    \begin{subfigure}[t]{0.27\textwidth}
        \centering
        \includegraphics[width=\textwidth]{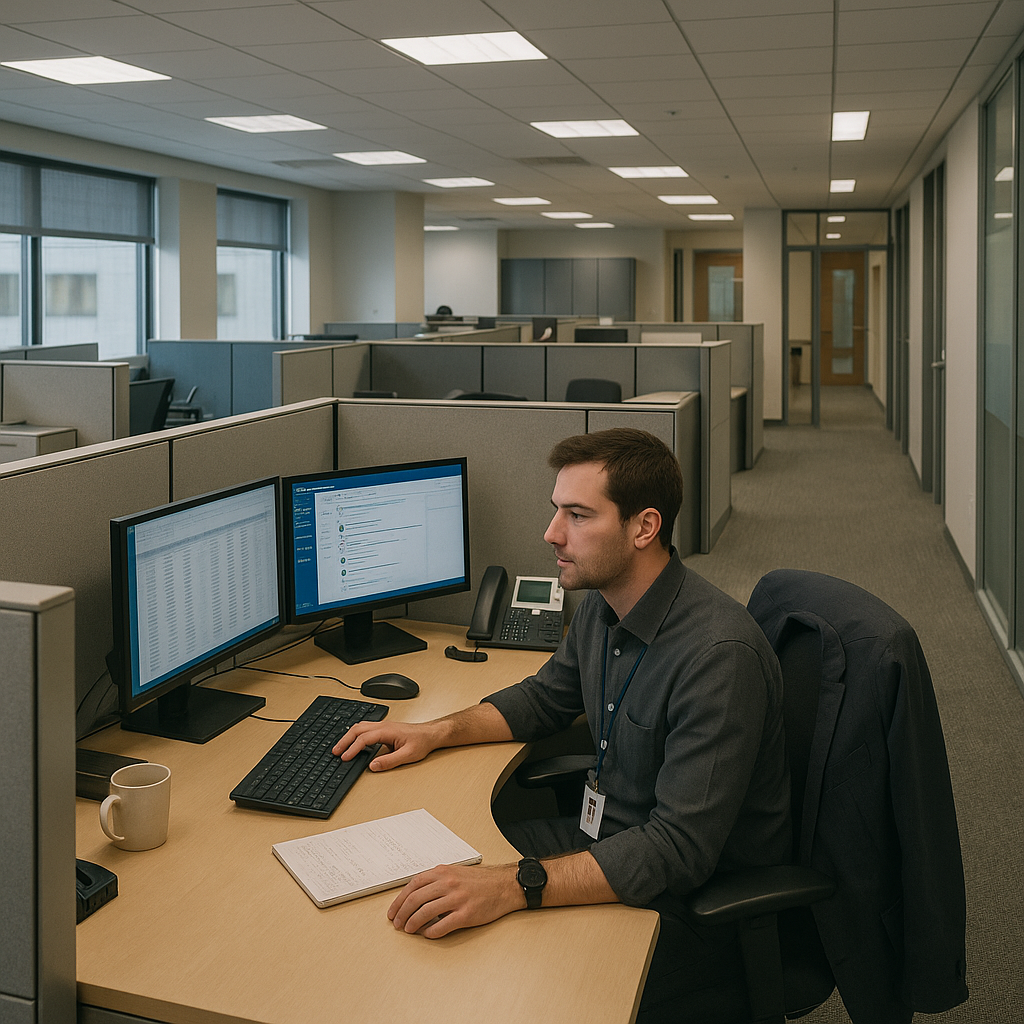}
        \caption{Office: Face only (base image).}
    \end{subfigure}
    \begin{subfigure}[t]{0.27\textwidth}
        \centering
        \includegraphics[width=\textwidth]{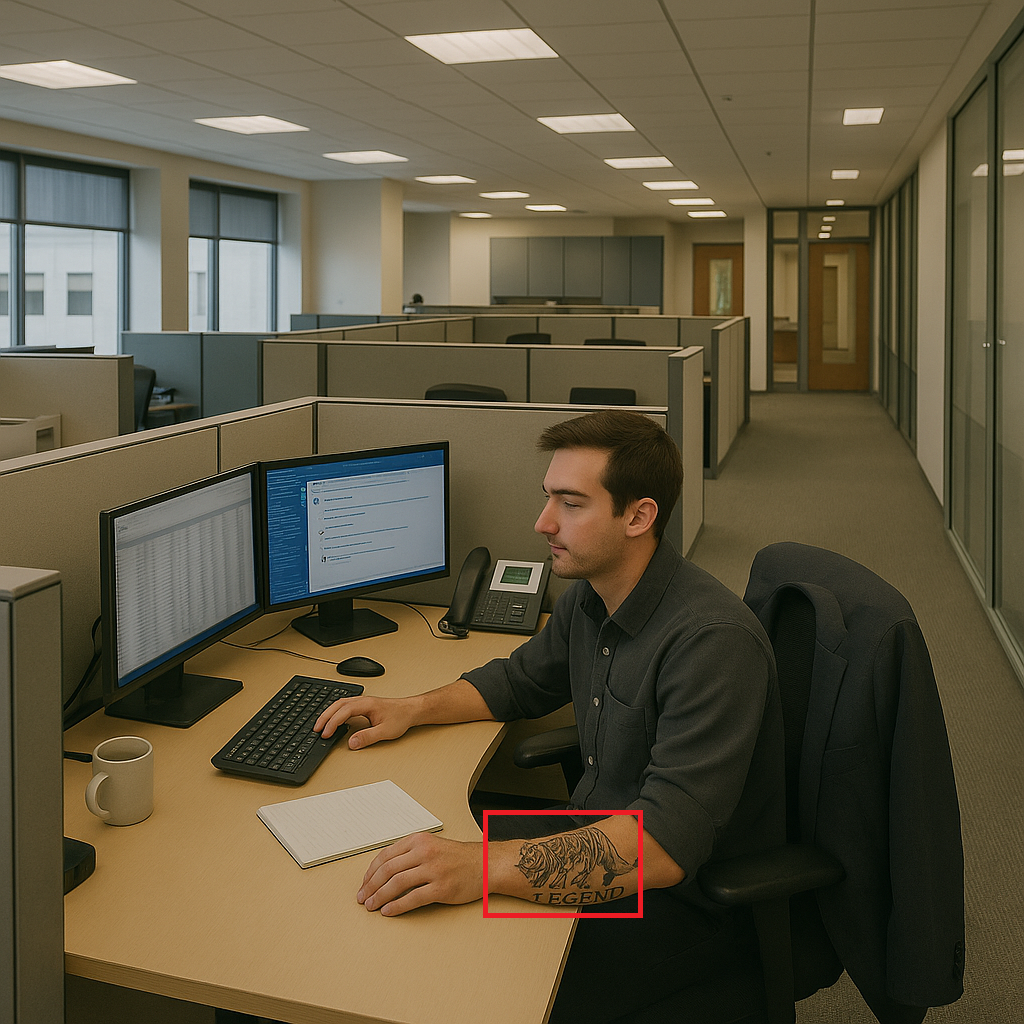}
        \caption{Face + Tattoo.}
    \end{subfigure}
    \begin{subfigure}[t]{0.27\textwidth}
        \centering
        \includegraphics[width=\textwidth]{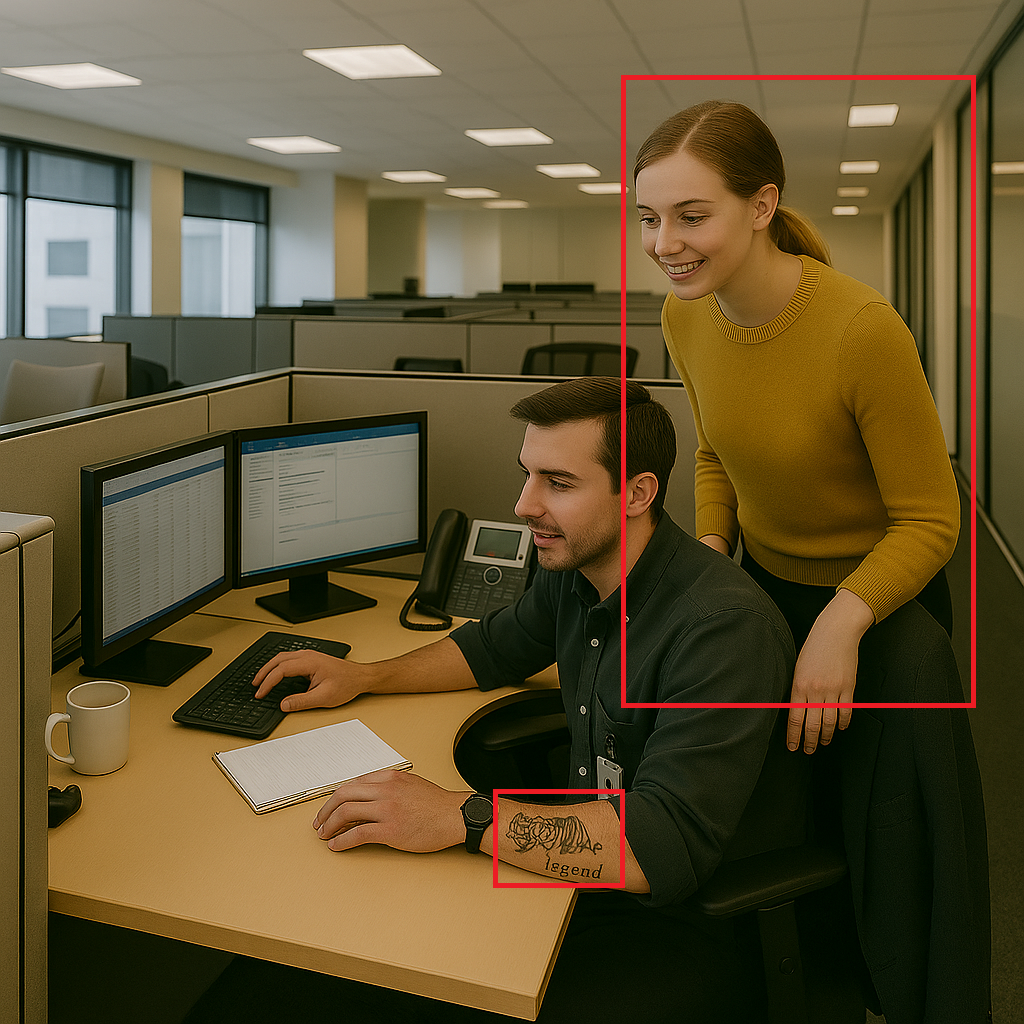}
        \caption{Face + Tattoo + Face (a.).}
    \end{subfigure}

    \caption{Illustrative examples of image generation in café (top) and office (bottom) environments. Each sequence progresses from a face only foreground PSO to combinations with additional background PSOs.}
    \label{fig:cafe_office_examples}
\end{figure*}

We used OpenAI's SORA text-to-image model~\footnote{https://sora.chatgpt.com/} to generate all images. For images containing only a foreground PSO, we designed comprehensive prompts to produce realistic and contextually
meaningful scenes. To maintain visual consistency, we applied
SORA's \textbf{Remix} functionality: instead of regenerating an entire image, we selected specific regions and instructed
the model to add the required background PSO within that context. This approach preserved lighting, perspective, and composition under all conditions. The generated images did not deviate significantly from each other, so the focus remains on PSOs. 
 We used a male subject for the office environment and a female subject for the café.
%
The gender assignments are random and are designed solely to provide variation across the environments. It does not  convey or reinforce any potential gendered association biases. 
Figure~\ref{fig:cafe_office_examples} shows examples from both environments. The top row starts with the base café image containing only the face foreground PSO, followed by versions where medicine and then landmark are added as background PSOs. The bottom row starts with the base office image, followed by versions where the tattoo and then an additional face (a.) are added as background PSOs. 
The prompts we used for image generation are provided in Appendix~\ref{app:prompts}.
%

For each foreground PSO, we generated images for three cases: (a) the foreground PSO alone ($1$ condition), 
(b) the foreground PSO individually paired with each of its five background PSOs ($5$ conditions), and 
(c) the foreground PSO combined with every possible pair of background PSOs ($\binom{5}{2} = 10$ conditions). 
This yields $1 + 5 + 10 = 16$ image conditions per foreground PSO. With three foreground PSOs, this corresponds to
$3 \times 16 = 48$ images per environment, which means that in total $48 \times 2 = 96$ images were generated for the two versions of the survey.

\subsubsection{Branch Logic Design}
Requiring participants to evaluate 48 images could lead to participant fatigue, reduced engagement, and unreliable responses due to rushed decision making. To mitigate this, we used branch logic to direct participants to different flows based on their responses to ranking comfortability levels when the image contains background PSOs. Instead of rating all possible combinations, participants first ranked the five background PSOs by sensitivity. 
Only the combinations involving the background PSO with the lowest effect on the foreground PSO's perceived privacy level (PPL) level were then displayed. As a result, participants rated $1 + 5 + 4 = 10$ images per foreground PSO and $30$ images in total. Each participant views identical $3\times(1+5)=18$ images from the same environment, while the last 12 differ due to branching. This approach also avoided redundant evaluation: If the PPL of a foreground PSO is highest when paired with a certain background PSO, then requesting participants to reassess it alongside other background PSOs becomes redundant, since the foreground PSO is already perceived as the most sensitive across all pairings. Appendix~\ref{appendix:full_survey_cafe} shows how the branch logic is implemented when the environment is café, the foreground PSO is face, and the landmark is chosen by the background PSO with the lowest effect. 




\subsection{Study Practicalities}
\label{subsec:organization}

\subsubsection{Survey Organization}
\label{sec:organization}

The first page of the survey contains the informed consent form, 
explaining the general objective of the study, how the data will be collected and processed, and compliance with GDPR. Proceeding the survey requires the confirmation of all mandatory consent statements, and an optional checkbox allows entry into a lottery for a movie ticket. Once consent is obtained, participants respond to demographic questions and questions about their social media habits and privacy awareness.
%
%
The privacy awareness questions included four brief items to understand individual privacy preferences in general. Following the approach of Hoyle et al.~\cite{hoyle2020privacynorms}, we opted for a short scale with minimal burden on respondents rather than a full psychometric scale such as the IUPUC~\cite{malhotra2004internet}.
%
Following this, participants complete three sections, each focusing on a different foreground PSO. 
In total, the survey contains 50 mandatory questions, except for the optional open-text fields. 
We should note that we do not evaluate whether individuals with stronger privacy preferences are more likely to perceive PSOs as private (i.e., less comfortable sharing them online). Our focus is on the \emph{effect of context} rather than individual privacy awareness. Moreover, previous work~\cite{hoyle2020privacynorms} has shown that higher privacy awareness is often associated with perceiving visual content, and attributes shape information sharing behaviors~\cite{Belanger2016Privacy}. Nevertheless, we acknowledge that the reported comfortability levels may vary depending on the privacy awareness of the participants. Individuals with stronger privacy concerns might rate PSOs as less comfortable to share, while those with lower awareness may perceive the same objects as less sensitive.

In our user study, each section begins with an image containing a single foreground PSO, and participants are asked about their comfortability level of sharing this object. The initial question is necessary to obtain a baseline perceived privacy level (PPL). The participants then view a row of images showing the same foreground PSO that is paired with a single background PSO different for each image, and are asked for the comfortability level for the same foreground PSO. They also rank the effect of background PSO on the degree of change and are asked about the rationale behind their ranking choice (optional open-text answer). The branch logic design is utilized according to their ranking. In the final stage, participants view a row of images including the same foreground PSOs, the background PSO ranked with the lowest decrease in their comfortability level (lowest PPL effect) paired with the remaining background PSOs. The final stage also includes an optional question asking to describe whether and in what way the lowest-ranked background PSOs influenced the comfortability level of the foreground PSO when combined with other background PSOs. 

All ratings used a 5-point Likert scale, where 1 indicated ``not comfortable at all'' and 5 ``very comfortable''. We asked for comfortability level instead of PPL since it is easier for participants to interpret while still naturally capturing the perceived privacy. Table~\ref{tab:comfort_privacy_mapping} in Appendix~\ref{appendix:additional_tables} shows the inverse mapping between the comfortability level and PPL. 
The complete survey questionnaire is provided in the Appendix~\ref{appendix:full_survey_cafe}.

\subsubsection{Survey Deployment}
%
Based on a careful review of the data storage policies, customizability, and control logic offered by different survey platforms, we selected Streamlit~\footnote{https://streamlit.io/} to implement the survey from scratch. We deployed the survey online and provided participants with a shared link to access it. Streamlit stores responses locally during completion and sends them to a designated email address upon submission. We downloaded these responses to a local machine for analysis.

\subsubsection{Participant Recruitment and Demographics}

Participants were recruited using a chain-referral (snowball) sampling approach. We initially distributed the survey link through our personal contacts and social media channels, reaching 73 individuals, and encouraged them to share it further within their own networks. In total, 109 participants completed the survey, with 61 assigned to the café version and 48 to the office version. After excluding responses that showed inconsistencies between ratings and rankings, 92 valid responses remained (51 café and 41 office). 
Participants were compensated with movie tickets through voluntary participation in a raffle.

Table~\ref{tab:demographics} shows the demographic distribution of the participants. The participants were balanced in sex (51\% female, 47\% male) and predominantly White/Caucasian (63\%) or Asian (22\%). Most of the participants' age was in the range $25 -34$ (35\%) or $45 - 54$ (32\%), with smaller groups in the ranges $18-24$ (20\%), $35 - 44$ (10\%) and $55+$ (4\%). Regarding education, the majority had a bachelor's degree (35\%) or a master's degree (32\%), while a quarter (26\%) had a doctorate. More than half of the participants were employed full-time (57\%), and students made up 30\%. Social media usage varied: 37\% reported sometimes, 36\% rarely, and only 2\% daily usage.

\subsection{Analysis overview}

\subsubsection{Quantitative Analysis}
\label{subsec:data_analysis}


To evaluate~\hyperref[rq:bg]{\bf H1}, we used the non-parametric Wilcoxon signed-rank test, since our design compares paired measurements of the same foreground PSO with and without background PSO (repeated measure with post intervention). The null hypothesis ~\hyperref[rq:bg]{\bf H1} is ``There is \textbf{no} difference between the perceived privacy level of a foreground PSO when it appears alone versus when accompanied by a background PSO''. To test~\hyperref[rq:env]{\bf H2}, we compared the comfortability scores between different environments using the non-parametric Mann-Whitney U test, as each participant rate the same foreground PSO only in one of the environments (café vs. office as independent variable). The null hypothesis for~\hyperref[rq:env]{\bf H2} is ``The distributions of perceived privacy levels are the same for café and office''. 
For~\hyperref[rq:co]{\bf H3}, we did not perform statistical testing, as the subset of participants who answered each combination question varied by individual rankings, making group-level inference unreliable (less number of repeated measures). Instead, we report the mean change in comfortability level and the proportion of participants who changed their comfortability level when two background PSOs were present. In addition these tests, we used the Mann-Whitney U test (with single categories) to understand whether demographics influence perceived privacy levels. To test factors with more than two categories, including age group, profession, and educational level, we used the Kruskal–Wallis H test.

\subsubsection{Qualitative Analysis}
\label{subsec:qualanalysis}
%
%
We conducted a qualitative analysis of the open-text responses from the survey questionnaire using inductive, in-vivo coding followed by thematic clustering to group related codes into higher-order themes~\cite{braun2023doing, blandford2016qualitative}. First, the author with an HCI background performed an independent review of the responses to get acquainted with the data using memoing and to generate the initial list of codes. 
Then, the author iteratively reviewed the initial codes along with the memos to deduce themes that could represent users' cognition or behaviour while dealing with visual privacy. This process was repeated until saturation was reached, with no further codes and themes or sub-themes emerging. Finally, the author drafted a codebook along with the definitions, inclusion, and exclusion criteria. The codebook was then reviewed to refine the wording of codes, their definitions, and thematic categories. The final codebook was reviewed once again to ensure that it resonated with the research questions and represented the survey data. Two authors then used the codebook to code all individual responses independently. 
Since the participants provided their responses as a summary of their cognitive walkthrough during our study, it is possible to tag the responses with multiple codes. Due to this overlap, the coders were allowed to use up to three codes per response. To assess reliability at the theme level, we collapsed the codes into their respective themes. We quantified agreement among coders using Krippendorff's alpha as the inter-rating reliability (IRR) measure due to its suitability ~\cite{mcdonald2019reliability}. We found the IRR to be 0.891, which indicates that coders consistently identified the same overarching themes and a strong agreement at the theme level, supporting the reliability of our thematic analysis.



\begin{table*}[t]
    \centering
    \caption{Responses to Privacy Awareness Questions (\%) and Mean Value. (SD = Strongly Disagree (1), SWD = Somewhat Disagree (2), N = Neutral (3), SWA = Somewhat Agree (4), SA = Strongly Agree (5)).}
    \label{tab:privacy_attitudes}
    \begin{tabular}{p{0.65\textwidth}cccccc}
    \toprule
    \textbf{Question} & \textbf{SD} & \textbf{SWD} & \textbf{N} & \textbf{SWA} & \textbf{SA} & \textbf{Mean} \\
    \midrule
    {Q1: I am concerned about my privacy online} & 1\% & 1\% & 8\% & 30\% & 60\% & 4.47 \\
    {Q2: I am concerned about my privacy in everyday life} & 2\% & 7\% & 14\% & 32\% & 46\% & 4.12 \\
    {Q3: It bothers me to give personal information to so many online companies} & 1\% & 1\% & 4\% & 30\% & 64\% & 4.52 \\
    {Q4: When I share a photo in social media, I check whether the picture contains personal information about me} & 8\% & 10\% & 8\% & 24\% & 50\% & 4.00 \\
    \bottomrule
    \end{tabular}
\end{table*}

\subsection{Study Ethics}
This study was conducted in accordance with regional privacy compliance and well-established best practice guidelines for user studies~\cite{redmiles2017summary,rea2014designing,egelman2007security}. Before the actual survey began, the participants were presented with information about the use of study results and the participation reward terms. The participation was voluntary, and we sought explicit consent from each participant who attempted the survey. The consent and privacy policy included in our study was drafted and provided by the data privacy regulation unit of our institution. A high-level privacy impact assessment was also conducted to review the study methodology and data to be collected to ensure ethical treatment of sensitive data. We ensured that personally identifiable information was solely used for contacting the participants and excluded from the analysis. We used an open-source, GDPR-compliant tool to build and host our study. All the data is entirely under our control on a server in the EU region, and it will be discarded upon the camera-ready version of the manuscript. The visual images used in our study were purely illustrative and did not depict any real human subjects. 


\section{Results}

Before analyzing the results, the responses that showed clear inconsistencies between the rankings and ratings were excluded from the dataset. Inconsistency refers to instances where a participant's assessment of the comfortability level concerning a foreground PSO, when there is a single background PSO, conflicts with the subsequent ranking question. This method also allowed us to filter out inattentive participants, thereby improving the quality of the remaining data.

\subsection{Effect of Users' Privacy Attitudes}


%

Table~\ref{tab:privacy_attitudes} presents the participants' responses to questions concerning privacy awareness. Concern about online privacy (Q1, 90\% agreed) and discomfort with the disclosure of personal information to online companies (Q3, 94\% agreed) received the highest ratings. Concern about privacy in daily life (Q2, 78\% agreed) was also significant but slightly lower overall. From the responses, we inferred that the majority of the participants expressed their concerns about their privacy online, in everyday life, and with the companies. 
The responses to (Q4) show that the participants' ratings of how often they check personal information before sharing a photo on social media were lower than their ratings of concern for privacy online.
This indicates that while most participants express concern about online privacy, this concern does not consistently translate into proactive behaviors, including looking for personal information before sharing it on social media, highlighting a gap between privacy attitudes and practices.

          

\subsection{Effect of Background PSOs (\hyperref[rq:bg]{H1})}



\begin{figure*}[t]
    \centering
    \begin{subfigure}[t]{0.30\textwidth}
        \centering
        \includegraphics[height=0.90\textwidth]{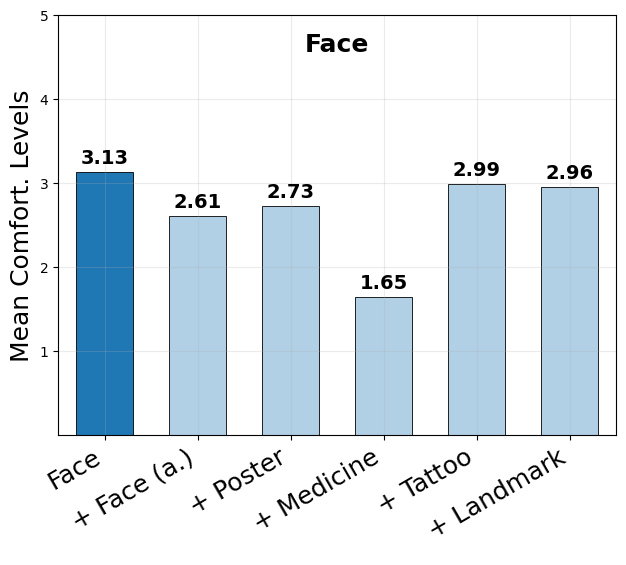}
        \caption{Face}
    \end{subfigure}
    \hfill
    \begin{subfigure}[t]{0.30\textwidth}
        \centering
        \includegraphics[height=0.90\textwidth]{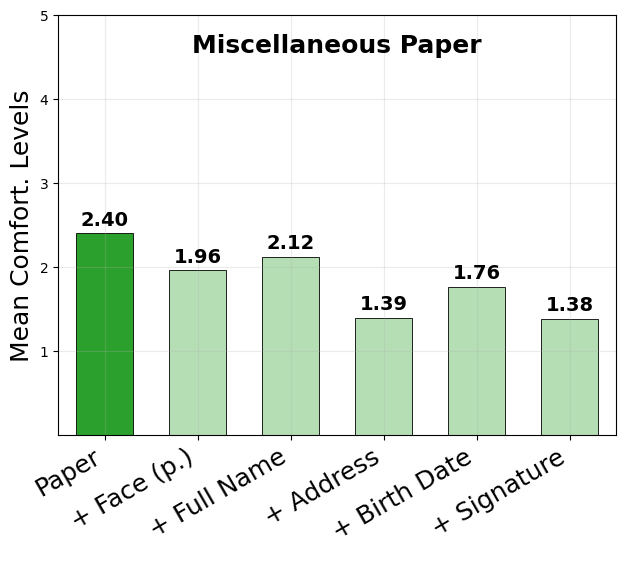}
        \caption{Miscellaneous Paper}
    \end{subfigure}
    \hfill
    \begin{subfigure}[t]{0.30\textwidth}
        \centering
        \includegraphics[height=0.90\textwidth]{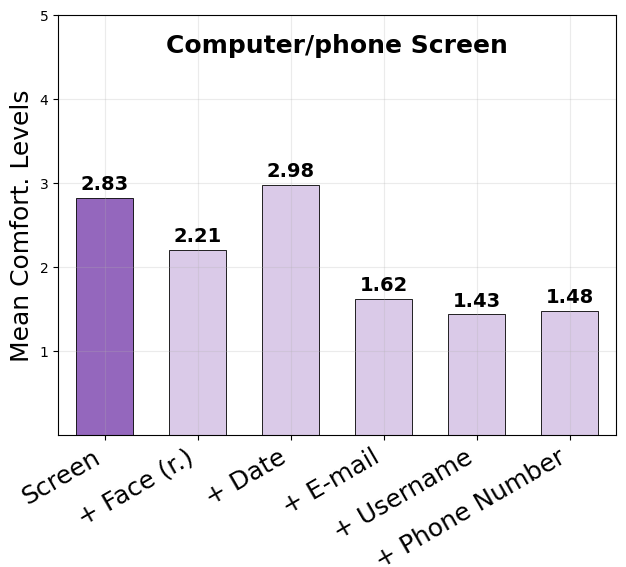}
        \caption{Computer/phone Screen}
    \end{subfigure}
    \caption{Average comfortability level ratings for foreground PSOs and the effect of a single background PSO.}
    \label{fig:bars}
\end{figure*}

To examine~\hyperref[rq:bg]{\textbf{H1}}, we quantitatively compared comfortability levels and averaged using both environments. Figure~\ref{fig:bars} 
illustrates how the presence of background PSO influences the comfortability level (the perceived privacy level, PPL) of each foreground PSO. In each figure, the first bar indicates the baseline comfortability level when the image contains only the foreground PSO, while the subsequent bars represent its combination with different background PSOs. For \textbf{Face}, the strongest reduction occurs with \emph{medicine} (mean = 1.65; –1.48 from the baseline of 3.13), followed by 
\emph{face (a.)} (2.61; –0.52). In contrast, \emph{tattoo} (2.99) and \emph{landmark} (2.96) produce minor decreases. For \textbf{Miscellaneous Paper}, \emph{full name} shows the smallest drop (2.12; $-0.28$), whereas \emph{address} (1.39) and \emph{signature} (1.38) yield the largest decreases. For \textbf{Computer/phone Screen}, identifiers such as \emph{email} (1.62), \emph{username} (1.43), and \emph{phone number} (1.48) lead to strong reductions, \emph{face (r.)} causes a moderate decrease (2.21). Interestingly \emph{date} stands out as the only case associated with an increase. A possible reason for this might be the averaging of rankings from both settings, which influences the participants' judgements. Overall, the addition of background PSOs 
mostly elevates the PPL of the foreground PSO, with certain categories (e.g., medicine for face, signature for miscellaneous paper, username for screen) exerting the most pronounced effects.

\begin{table*}[t]
\centering
\caption{Wilcoxon signed-rank test results: Change in comfortability level ($\Delta$ mean) when adding a background PSO. Significant $p$-values ($<0.05$) are highlighted in \textcolor{blue}{\textbf{bold and blue}}.}
\label{tab:bg_results}
\begin{adjustbox}{max width=\textwidth}
\begin{tabular}{c c c c c}
\toprule
\textbf{Foreground PSO} & \textbf{Background PSO} & \textbf{Café ($\Delta$, $p$)} & \textbf{Office ($\Delta$, $p$)} & \textbf{Significance} \\
\midrule
\multirow{5}{*}{Face}    & Face (a.)      & -0.37 (0.0576)          & -0.71 (\textcolor{blue}{\textbf{0.0003}}) & Only Office \\
        & Poster    & -0.55 (\textcolor{blue}{\textbf{0.0345}}) & -0.22 (0.2193)          & Only Café \\
        & Medicine  & -1.78 (\textcolor{blue}{\textbf{0.0000}}) & -1.10 (\textcolor{blue}{\textbf{0.0000}}) & Both \\
        & Tattoo    & -0.22 (0.4359)          & -0.05 (0.6917)          & None \\
        & Landmark  & -0.16 (0.4115)          & -0.20 (0.1895)          & None \\
\midrule
\multirow{5}{*}{Misc. Paper}     & Face (p.)  & -0.33 (0.0815)    & -0.59 (\textcolor{blue}{\textbf{0.0049}}) & Only Office \\
                & Full Name  & -0.25 (0.1256)    & -0.32 (0.0920)          & None \\
                & Address    & -1.22 (\textcolor{blue}{\textbf{0.0000}}) & -0.76 (\textcolor{blue}{\textbf{0.0006}}) & Both \\
                & Birth Date  & -0.57 (\textcolor{blue}{\textbf{0.0039}}) & -0.73 (\textcolor{blue}{\textbf{0.0012}}) & Both \\    
                & Signature  & -1.16 (\textcolor{blue}{\textbf{0.0000}}) & -0.85 (\textcolor{blue}{\textbf{0.0001}}) & Both \\
\midrule
\multirow{5}{*}{Computer/phone Screen}   & Face (r.)       & -0.47 (\textcolor{blue}{\textbf{0.0165}}) & -0.80 (\textcolor{blue}{\textbf{0.0001}}) & Both \\
                & Date       & +0.25 (0.0849)          & +0.02 (0.7181)          & None \\
                & E-mail      & -1.37 (\textcolor{blue}{\textbf{0.0000}}) & -1.00 (\textcolor{blue}{\textbf{0.0005}}) & Both \\
             & Username   & -1.57 (\textcolor{blue}{\textbf{0.0000}}) & -1.17 (\textcolor{blue}{\textbf{0.0000}}) & Both \\
                & Phone Number     & -1.59 (\textcolor{blue}{\textbf{0.0000}}) & -1.05 (\textcolor{blue}{\textbf{0.0001}}) & Both \\
\bottomrule
\end{tabular}
\end{adjustbox}
\end{table*}

Figure~\ref{fig:bars} shows only mean values and does not capture the underlying distributions. To systematically examine whether observed differences were consistent, we compared the comfortability levels of each foreground PSO alone with their corresponding foreground/background PSO pair using the Wilcoxon signed-rank test. 
Table~\ref{tab:bg_results} shows that the presence of background PSOs often reduced the comfortability level, indicating an increase in perceived privacy level (PPL). For the \textbf{Face}, the addition of \textit{face (a.)} significantly increased PPL in the office but not in the café, suggesting that the effect of background PSO matters more in professional contexts. \textit{Medicine} strongly increased PPL in both environments, while \textit{tattoo} and \textit{landmark} had no significant effect. 
For \textbf{Miscellaneous Paper}, \textit{address, birth date}, and \textit{signature} consistently raised PPL across both environments. \textit{Face (p)} increased PPL only in office, while \textit{full name} had no significant effect. 
For \textbf{Computer/phone Screen}, nearly all background PSOs increased PPL, except \textit{date}, which has no effect.

Our analysis shows that \textbf{not all background PSOs equally affect perceived privacy}. Highly sensitive cues (e.g., medicine, addresses, signatures, digital identifiers on screens) consistently amplify the PPL of foreground PSOs, while more ambiguous cues (e.g., tattoos, landmarks, dates) are often ignored. This pattern reflects with everyday social media behavior, where users avoid sharing explicit sensitive information such as medicine or ID cards, but tend to underestimate the risks posed by ambigious cues that can be exploited in inference attacks such as social engineering or identity linking.

\subsection{Effect of Environment (\hyperref[rq:env]{H2})}

To evaluate~\hyperref[rq:env]{H2}, we first compared the average comfortability levels between environments. The comfortability level was consistently lower in office settings than in café, with the strongest drop observed for \textbf{Face} as a foreground PSO (Face: $ -0.67 $; Misc. Paper: $ -0.37 $; Screen: $ -0.31 $). To systematically assess these differences, we applied the Mann–Whitney U test. As shown in Table~\ref{tab:environment_h2}, participants reported significantly lower comfortability (i.e., higher perceived privacy level, PPL) in office settings across multiple foreground/background PSO pairs. The effect was especially pronounced for pairs that involve the face as a single foreground PSO ($p \!=\! 0.0062$), face/face(a.) ($p\!<\!0.001$), face/tattoo ($p\!=\!0.0261$), and face/landmark ($p\!=\!0.0032$); where office settings consistently amplified PPL. In contrast, paper-based PSOs (e.g., birthdate, face(p.)) yielded lower overall comfortability levels but smaller differences between environments. The violin plots in Appendix Figure~\ref{fig:sig_env} strengthen the analysis of~\hyperref[rq:env]{H2} by visually illustrating the underlying distributional patterns. Across all eight statistically significant pairings, participants consistently reported lower comfortability level in the \textbf{office} setting compared to \textbf{café}, reaffirming the context sensitivity of perceived privacy. Figure~\ref{fig:sig_env} further reveals changes in distribution spread, indicating a greater consensus on discomfort in office environments.

\begin{table*}[t]
\centering
\caption{Mann-Whitney U test results comparing comfortability levels between café and office environments. Positive $\Delta$Mean values indicate higher average comfortability in the café, while negative values indicate higher comfortability in the office. Significant $p$-values are highlighted in \textcolor{blue}{\textbf{bold and blue}}.}
\label{tab:environment_h2}

\small
\label{tab:env_results}
\begin{tabular}{ccc}
\toprule
\textbf{Face} & \textbf{Miscellaneous Paper} & \textbf{Computer/phone Screen} \\
\begin{tabular}{lcc}
\toprule
Background PSO & $\Delta$Mean & $p$-value \\
\midrule
None   & +0.81 & \textcolor{blue}{\textbf{0.0062}} \\
Face (a.)     & +1.14 & \textcolor{blue}{\textbf{0.0000}} \\
Poster    & +0.48 & 0.0518 \\
Medicine  & +0.12 & 0.4519 \\
Tattoo    & +0.64 & \textcolor{blue}{\textbf{0.0261}} \\
Landmark  & +0.85 & \textcolor{blue}{\textbf{0.0032}} \\
\end{tabular}
&
\begin{tabular}{lcc}
\toprule
Background PSO & $\Delta$Mean & $p$-value \\
\midrule
None    & +0.42 & 0.1193 \\
Face (p.)      & +0.67 & \textcolor{blue}{\textbf{0.0152}} \\
Full Name       & +0.48 & 0.1549 \\
Address    & -0.04 & 0.9488 \\
Birth Date  & +0.58 & \textcolor{blue}{\textbf{0.0043}} \\
Signature  & +0.11 & 0.2426 \\
\end{tabular}
&
\begin{tabular}{lcc}
\toprule
Background PSO & $\Delta$Mean & $p$-value \\
\midrule
None   & +0.43 & 0.0722 \\
Face (r.)      & +0.77 & \textcolor{blue}{\textbf{0.0098}} \\
Date      & +0.66 & \textcolor{blue}{\textbf{0.0057}} \\
Email     & +0.06 & 0.7789 \\
Username  & +0.04 & 0.7930 \\
Phone     & -0.11 & 0.4312 \\
\end{tabular}
\\
\bottomrule
\end{tabular}
\end{table*}


\begin{figure*}[t]
    \centering
    \begin{subfigure}[t]{0.30\textwidth}
        \centering
        \includegraphics[height=0.75\textwidth]{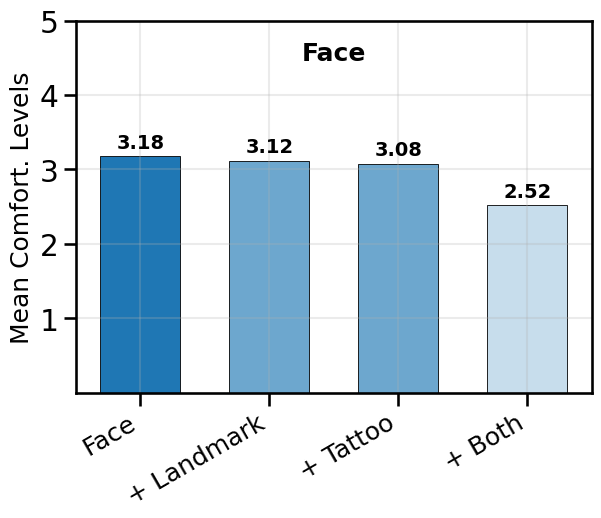}
        \caption{Face}
    \end{subfigure}
    \hfill
    \begin{subfigure}[t]{0.30\textwidth}
        \centering
        \includegraphics[height=0.75\textwidth]{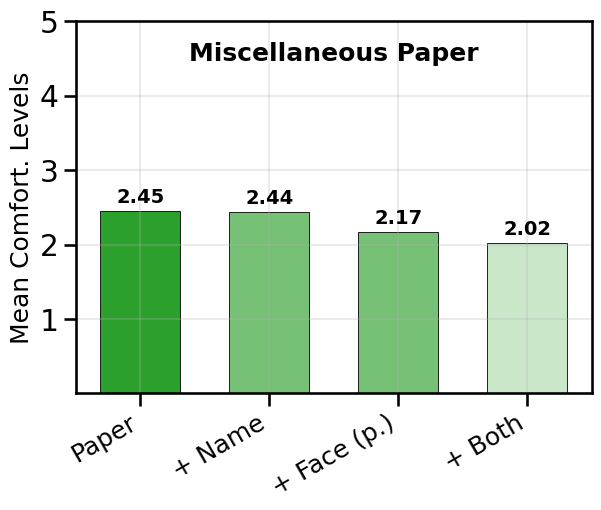}
        \caption{Miscellaneous Paper}
    \end{subfigure}
    \hfill
    \begin{subfigure}[t]{0.30\textwidth}
        \centering
        \includegraphics[height=0.75\textwidth]{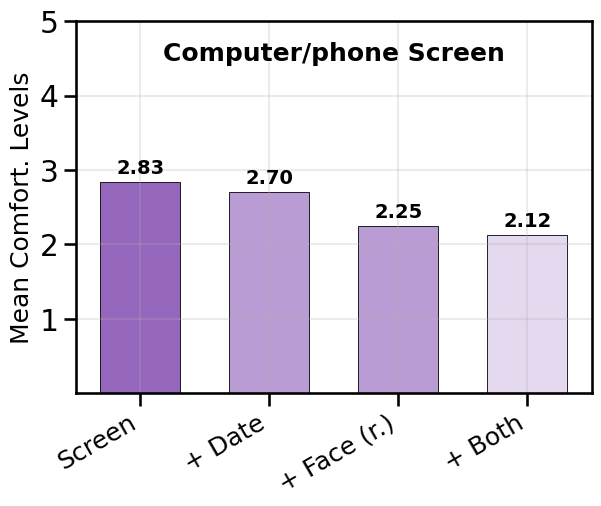}
        \caption{Computer/phone Screen}
    \end{subfigure}
    \caption{Average comfortability level ratings for each foreground PSO, showing the baseline, the effect of the two background PSO most often rated as having the lowest effect when alone (see Table~\ref{tab:least_private_counts}), and the effect of co-presence with both background PSOs present in the image (\textbf{both}).}
    \label{fig:cobars}
\end{figure*}

\begin{table*}[t]
\centering
\caption{Background PSOs rated as having the least effect on comfortability level for each foreground PSO. Values indicate counts with percentages for Café ($n=51$) and Office ($n=41$). For each environment and foreground PSO, the background PSO most frequently selected is highlighted in \textcolor{blue}{\textbf{bold and blue}}.}
\label{tab:least_private_counts}
\begin{adjustbox}{width=\textwidth}
\renewcommand{\arraystretch}{1.1}
\begin{tabularx}{\textwidth}{@{\extracolsep{\fill}} X c | X c | X c @{}}
\toprule
\textbf{Face} & Café / Office & \textbf{Misc. Paper} & Café / Office & \textbf{Screen} & Café / Office \\
\midrule
Face (a.)         & 13 (26\%) / 8 (20\%)  &  Face (p.)  & 15 (29\%) / 6 (15\%) &  Face (r.)   & 13 (26\%) / 4 (10\%) \\    
Poster            & 9 (18\%) / 10 (24\%)  & Full Name & 12 (24\%) / \textcolor{blue}{\textbf{19 (46\%)}} & Date  & \textcolor{blue}{\textbf{36 (71\%)} / \textbf{32 (78\%)}} \\
Medicine          & -- / 2 (5\%)  &  Address    & 4 (8\%) / 2 (5\%) & E-mail      & -- / 1 (2\%) \\
Tattoo            & 11 (22\%) / \textcolor{blue}{\textbf{10 (24\%)}} &  Birth Date  & \textcolor{blue}{\textbf{16 (31\%)}} / 6 (15\%) & Username   & 2 (4\%) / 2 (5\%) \\
Landmark & \textcolor{blue}{\textbf{18 (35\%)}} / 11 (27\%) & Signature  & 4 (8\%) / 8 (20\%) & Phone Number    & -- / 2 (5\%) \\
\bottomrule
\end{tabularx}
\end{adjustbox}
\end{table*}

These results suggest that not all PPLs of PSOs are equally shaped by the environmental context. Highly identifiable cues, such as faces, are perceived particularly private in professional settings, where co-presence may further heighten concerns. In contrast, more document-like PSOs show consistently low comfortability regardless of the environment. The violin plots add nuance to this interpretation by revealing how comfortability levels are lower in average in office settings, and more tightly distributed, indicating stronger agreement among participants. In conclusion, the context amplifies perceived privacy levels, and sensitivity is affected not only by single objects but also by their combinations and settings. This strengthens the claim that privacy-aware systems must move beyond binary recognition to incorporate compositional object definitions and contextual awareness. 


\subsection{Effect of Co-Presence (\hyperref[rq:co]{H3})}

\begin{table*}[t]
\footnotesize 
\centering
\caption{Co-presence effects of background/background PSO pairs on perceived privacy levels for the foreground PSO (Face, Misc. Paper, Screen). $n$ denotes the number of participants who rated the comfortability levels for a given pair. The \% Decrease column shows the proportion of these participants who rated a lower comfortability level than images where each  background PSO is shown separately, highlighting cases where co-presence amplifies privacy concerns.}

\label{tab:pairs_positive}
\begin{tabular}{ccc}
\toprule
\textbf{Face} & \textbf{Misc. Paper} & \textbf{Screen} \\
\begin{tabular}{lcc}
\toprule
Pair & n & \% Decrease \\
\midrule
poster/tattoo    & 40 & 37.5 \\
landmark/tattoo  & 50 & 28.0 \\
face (a.)/poster     & 40 & 22.5 \\
landmark/poster  & 48 & 20.8 \\
face (a.)/landmark   & 50 & 20.0 \\
landmark/medicine& 31 & 19.4 \\
medicine/poster  & 21 & 19.0 \\
medicine/tattoo  & 23 & 17.4 \\
face (a.)/tattoo     & 42 & 16.7 \\
face (a.)/medicine   & 23 & 13.0 \\
\end{tabular}
&
\begin{tabular}{lcc}
\toprule
Pair & n & \% Decrease \\
\midrule
full name/face (p.)       & 52 & 17.3 \\
birthdate/face (p.)  & 43 & 16.3 \\
address/full name     & 37 & 13.5 \\
birthdate/full name   & 53 & 13.2 \\
birthdate/signature & 34 & 11.8 \\
full name/signature   & 43 & 11.6 \\
address/birthdate& 28 & 10.7 \\
photo/signature  & 33 &  9.1 \\
address/face (p.)    & 27 &  7.4 \\
address/signature& 18 &  5.6 \\
\end{tabular}
&
\begin{tabular}{lcc}
\toprule
Pair & n & \% Decrease \\
\midrule
e-mail/username   &  5 & 60.0 \\
phone num./username   &  6 & 33.3 \\
face/username    & 21 & 23.8 \\
date/e.mail       & 69 & 18.8 \\
face/phone num.       & 19 & 15.8 \\
date/face (r.)        & 85 & 14.1 \\
e-mail/face (r.)       & 18 & 11.1 \\
date/phone num.       & 70 &  8.6 \\
date/username    & 72 &  2.8 \\
e-mail/phone num.      &  3 &  0.0 \\
\end{tabular}
\\
\bottomrule
\end{tabular}
\end{table*}

\begin{table*}[t]
\centering
\caption{Dominance effects of background PSOs on foreground PSOs (Face, Misc. Paper, Screen) during co-presence with other background PSOs.
Columns: \textbf{Background PSO} = background privacy sensitive object co-present with other background PSOs; \textbf{N} = number of co-presence ratings; \textbf{D} = number of dominance cases; \textbf{\%D} = proportion of dominance cases (D/N).}
\label{tab:bg_dominance}
\begin{tabular}{ccc}
\toprule
\textbf{Face} & \textbf{Misc. Paper} & \textbf{Screen} \\
\begin{tabular}{lccc}
\toprule
Background PSO & N & D & \%D \\
\midrule
Medicine  &  98 & 64 & 65.3 \\
Poster    & 149 & 40 & 26.8 \\
Face (a.)     & 155 & 38 & 24.5 \\
Landmark  & 179 & 39 & 21.8 \\
Tattoo    & 155 & 27 & 17.4 \\
\end{tabular}
&
\begin{tabular}{lccc}
\toprule
Background PSO & N & D & \%D \\
\midrule
Address   & 110 & 36 & 32.7 \\
Signature & 128 & 41 & 32.0 \\
Face (p.)     & 155 & 25 & 16.1 \\
Birth Date & 158 & 24 & 15.2 \\
Full Name      & 185 & 20 & 10.8 \\
\end{tabular}
&
\begin{tabular}{lccc}
\toprule
Background PSO & N & D & \%D \\
\midrule
Username & 104 & 61 & 58.7 \\
Phone num.   &  98 & 52 & 53.1 \\
E-mail    &  95 & 45 & 47.4 \\
Face (r.)     & 143 & 35 & 24.5 \\
Date     & 296 & 17 &  5.7 \\

\end{tabular}
\\
\bottomrule
\end{tabular}
\end{table*}

We evaluated~\hyperref[rq:co]{H3} by testing whether the co-presence of background PSOs further increases the perceived privacy level (PPL) of a foreground PSO compared to the presence of a single background PSO. 
To avoid survey fatigue from rating all possible combinations, we adopted a branching procedure: each participant first identified the background PSO that had the lowest effect on a given foreground PSO, and this ``least effective'' background PSO was paired with each remaining background PSO for the same foreground PSO. Table~\ref{tab:least_private_counts} reports how often each background PSO was assessed as the least effective, determining the combinations evaluated. The resulting selection imbalance made formal statistical testing approaches inappropriate. Therefore, we report two complementary quantitative measures: (1) the mean change in the comfortability level when a second background PSO is added to the image and (2) the proportion of participants whose comfortability level decreased after this addition. 

Figure~\ref{fig:cobars} illustrates the magnitude of mean changes when two least effective background PSOs (extracted from Table~\ref{tab:least_private_counts}) are present in the image, labeled ``both''. Table~\ref{tab:pairs_positive} reports the proportion of participants who decreased their level of comfortability when they observe two background PSOs. Both results converged to the same pattern: a small subset of PSO pairs consistently decreased the comfortability level (amplified perceived privacy level) beyond their single-object effects. For example, the co-presence of landmark/tattoo produced a considerable average drop in comfortability level ($0.66$) and in a substantial fraction of the participants, the co-presence reduced the willingness to share the foreground PSO. Pairs such as poster/tattoo and landmark/tattoo repeatedly ranked among the most impactful, with other notable combinations including e-mail/username and phone number/username on computer/phone screen. Most other co-presences showed minimal effects.

To further explore co-presence, we examined whether certain background PSO exerts dominance effect, meaning that when they appeared together with another background PSO, participant's comfortability levels aligned almost exclusively with the more influential cue. Table~\ref{tab:bg_dominance} summarizes these results. For the \textbf{Face} foreground PSO, \textit{medicine} dominated almost two-thirds of the cases (65.3\%), indicating that the participants largely ignored the other background PSO where medicine was present. Similarly, for \textbf{Misc. Paper}, \textit{address} (32.7\%) and \textit{signature} (32.0\%) often dominated the influence of other background PSOs. For the \textbf{Screen}, identifiers such as \textit{username} (58.7\%), \textit{phone number} (53.1\%), and \textit{e-mail} (47.4\%) showed strong dominance, while \textit{date} (5.7\%) rarely dominated other background PSOs. 
Importantly, lower $N$ values indicate that more participants initially considered these PSOs to be relatively more effective on their own, making their strong dominance effects even more notable. For example, \textit{medicine} was often chosen as the most effective background PSO when in isolation and overwhelmingly dominated the participant's judgment when paired with others.

Our findings show that some PSOs act as \textbf{primary drivers of perceived privacy risk}, overshadowing other co-present cues. This suggests that privacy-aware systems should prioritize flagging dominant cues instead of treating every cue equally, as these alone often drive users' judgments, making fine-grained flagging of weaker cues less critical for a better usability.

\bolditalictitle{Demographic Effects:}
We also examined whether demographic factors influenced privacy perceptions. Significant effects are summarized in Table~\ref{tab:sig_demographics} in Appendix~\ref{appendix:additional_tables}. In general, gender effects were the most present, with female participants reporting higher comfortability levels, while other demographic factors showed more isolated differences. Our results suggest that privacy-aware systems may benefit from considering demographic variability, particularly when tailoring defaults for awareness mechanisms.

\subsection{Qualitative Insights}
\label{subsec:qualins}

Qualitative thematic analysis (refer to Section~\ref{subsec:qualanalysis}) of responses to open-ended survey questions yielded ten codes spanning across three discrete themes. The refined codebook is shown in Table~\ref{tab:themes}. 
Our results indicate that participants have a latent cognitive model when self-evaluating the privacy of visual content that comprises multiple sensitive elements. In particular, the identified themes depict a layered model with heuristics, deeper reasoning, and consequence weighing as follows: First, the participants have mental shortcuts and norms (such as ``\textit{Everyone does it this way; So, it is fine!}'') as comfort heuristics. If such heuristics raise a red flag, they dig deeper into inference (such as ``\textit{What if someone figures out more?}'') and potential consequences (such as ``\textit{Would that be dangerous?}'').

\begin{table}[t]
\centering
\caption{Thematic analysis results.}
\renewcommand{\arraystretch}{1.3} 
\setlength{\tabcolsep}{5pt} 
\begin{tabular}{|m{0.15\linewidth} |m{0.20\linewidth} | m{0.50\linewidth} |}
\hline
\textbf{Theme} & \textbf{Code} & \textbf{Definition} \\
\hline

\multirow{5}{*}{\shortstack[1] {Perceived \\Privacy \\Norms}}   
& Social (media) norms & Participants feel safer sharing when attention is distributed or others are present. \\
\cline{2-3}
& Need for consent & Participants seek permission before posting others to respect privacy and consent.  \\
\cline{2-3}
& Tolerance for incremental exposure & Disclosure risk is perceived as low when similar information is already public, reducing reluctance to incrementally share more.  \\
\hline

\multirow{6}{*}{\shortstack[1] {Inference \& \\ Linkability}} 
& Contamination risks & Concerns about being associated with unrelated or unwanted ideological affiliations that could create false reputation.  \\
\cline{2-3}
& Spatio-temporal pinpointing & Concerns about the combination of location and time information from visual content.  \\
\cline{2-3}
& Temporal unlinkability & Concerns about time-based linkage or exposure. \\
\cline{2-3}
& Aggregation risks & Risk multiplies when multiple identifiers combine, enabling identity theft or profiling even if each item alone seems safe.  \\
\hline

\multirow{3}{*}{\shortstack[1] {Irreversible \\ Harms}} 
& Uniqueness & Concerns about unique features affecting personal privacy. \\
\cline{2-3}
& Stigmas & Certain sensitive information is perceived as categorically too risky, as their exposure leads to stigma, discrimination, or fraud. \\
\cline{2-3}
& Threat actors & Concerns about potential adversaries and their capabilities. \\
\hline
\end{tabular}

\label{tab:themes}
\end{table}

\bolditalictitle{Perceived Privacy Norms:} (See quotes Q1 and Q2) This theme explores the comfort heuristics inferred by participants based on their interpretation of social expectations, consent norms, and perceived acceptability while sharing photos online. 

\begin{quotebox2}
{\small \textbf{Q1:} ........... if there's someone else in the photo I might feel more comfortable sharing rather than being just me. ...........} 
\end{quotebox2}

This theme's codes included instances where the participants mentioned the co-presence of other individuals and situations requiring mutual consent for the photo to be shared online. The codes also covered scenarios where the public availability of sensitive information reduced privacy concerns for revealing additional data. We excluded quotes related to non-human objects. 

\begin{quotebox2}
{\small \textbf{Q2:} .......... I usually post other humans when it is some social event, like a party or celebration, so by default everyone is aware pictures are taken and they usually say if they don't like them so they don't want them posted ........... }
\end{quotebox2}

\bolditalictitle{Inference and Linkability:} (See quotes Q3 and Q4) This theme captures a deeper reasoning by a user about how information in an image can reveal hidden attributes or be linked across datasets to identify, profile, or de-anonymize themselves or others present in the shared image. 

\begin{quotebox2}
{\small \textbf{Q3:}..... Poster can contain certain propaganda that I do not affiliate myself with, so I wouldn't want it showing up. ......} 
\end{quotebox2}

The underlying codes of this theme capture scenarios of risks related to being associated with unrelated topics or the combination of multiple PII in the photo that reveal more than intended. Furthermore, they also included concerns about the disclosure of real-time locations directly through the cues in the shared photo or indirectly through the metadata. Some codes also reflect ad-hoc strategies employed by participants to mitigate the aforementioned risks.

\begin{quotebox2}
{\small \textbf{Q4:}.......... A combination of my (full) name plus all the other would be considered an information breach for me and I would not feel comfortable of people that I am not close with (as is the case in social media) knowing all this stuff about me plus it would create a high risk ........... }
\end{quotebox2}

\bolditalictitle{Irreversible Harms:} (See quote Q5) This theme captures specific visual objects whose disclosure can cause long-lasting harms that cannot be easily remediated. The codes reflect instances where the participant expresses concerns about privacy risks associated with disclosure of, for example, distinctive physical attributes (e.g., tattoos or birthmarks) or health conditions. The codes also covered responses that mention well-known cyber-threat actors and the permanent consequences they could cause. General privacy concerns and non-sensitive information leaks were excluded from this theme.

\begin{quotebox2}
{\small \textbf{Q5:}..... Address is also very personal, especially to females, due to possibilities  of stalkers, kidnappers, or robbers. Birthdate can be used for some type of fraud as well, given that in many countries the personal code contains birthday date......} 
\end{quotebox2}

In summary, the results of the qualitative analysis emphasize various aspects of user mental models for visual privacy self-evaluation. The \textit{Perceived Privacy Norms} theme reflects ad-hoc strategies for managing comfort. \textit{Inference and Linkability} captures the risk assessment. The \textit{Irreversibility Harms} theme reflects on the potential dangers if the risk is actualized.








\section{Discussion}
One of our key observations is that users' privacy perceptions do not always translate into privacy behavior or user actions. For example, the results from the privacy awareness questions (Table~\ref{tab:privacy_attitudes}) show that the participants expressed general concern about online privacy, but not all examined the photos for personal information before sharing. Qualitative data from the survey also displays several instances of privacy awareness among participants, which appears to have been formed through public news media debates around privacy concerns and mandatory training at the workplace and educational institutions. Despite the awareness, users often fail to assess the privacy risks or control their behavior while sharing visual content on online platforms~\cite{cheung2016evaluating,barnes2006privacy,gerber2018explaining}. The lack of proactive practices highlights that current systems fail to bridge the gap between users' mental models and actions.

The complexity of privacy–utility tradeoffs further contributes to this gap. Usually, foreground PSOs carry inherent sensitivity. However, people may choose to share them because of the tangible benefits they provide, such as identity verification, professional collaboration, or social connection. Our results show that the perceived privacy risks associated with these PSOs are not fixed but dynamically reshaped by the presence of background PSOs. For example, a selfie becomes more sensitive when medicine is visible, a document becomes more risky to share when a signature is present, and a screenshot becomes problematic when the username or phone number is exposed. Such contextual amplification indicates that privacy–utility tradeoffs, especially in the case of visual contents, are a rather complicated process beyond a simple binary choice. The complexity in privacy decision-making arises from users' attempts to weigh multiple contextual factors and engage in multidimensional self-negotiation for the tradeoffs.  

These observations emphasize the need for novel technical interventions and research opportunities to explore the notion of context-aware privacy. In this realm, the hypothesis evaluation and results from our study provide insight into the design principles and required characteristics of a context-aware privacy system. From a system design point of view, such a system should avoid uniform treatment of visual elements and adapt privacy sensitivity based on contextual cues. 
%
Also, the design should consider two broad categories of contextual cues: (i) subtle, latent cues (e.g., religious symbols or political affiliations) that are hard to detect, and (ii) categorically sensitive elements (e.g., medicine) that have a dominance effect by overriding other cues disproportionately influencing user perceptions, as users often base their privacy judgments on such cues instead of the full context.
Moreover, we observed that the co-presence of certain visual elements can lead to even higher perceived privacy risks. 
Thus, we argue that privacy-aware system design benefits from integrating object-level detection with contextual inference that accounts for the dominance and co-presence of visual elements.


From an HCI design perspective, privacy-aware systems should provide automated assistance and fine-grained control to users. The assistance can be provided in the form of nudges, via visual highlighting and prompts, that guide users towards low-burden, privacy-oriented actions. Alternatively, assistance can also be provided through the automatic detection of highly sensitive objects or contextual cues in the background that could amplify the perception of privacy. 
%
%
 Automatic assistance should also suggest appropriate privacy controls, allowing users to selectively obfuscate specific sensitive objects of the image while retaining the utility of the image and preserving privacy in its redacted form.
Within this context, we argue that a human-centered design approach can assist users in translating their privacy perceptions into concrete actions, reducing cognitive load and improving decision-making.

Existing online social media and instant communication platforms, which deal with user-generated visual content, largely lack user assistance and nuanced controls for privacy. 
The privacy settings for the user accounts offered by some of these platforms fail to address the risks highlighted in our work. The results and discussions of our work can provide actionable insights to integrate visual privacy protection components. These insights can also guide usable system design themes of research works and development of future technologies that deliver personalized privacy protection.

\subsection{Limitations and Future Work}

This subsection discusses the limitations of our work in terms of methodological trade-offs and factors affecting generalizability, which may constrain the external validity of our results, as well as outline future research directions.

Most of the participants in our study identify themselves as White/Caucasian, highly educated, younger and mid-career adults. This skewed demographic sample may limit the generalizability of our results regarding cultural and personal background diversity, as such factors shape users' privacy awareness and perceptions. Likewise, although we used a representative set of privacy-sensitive objects, this finite set may not capture the full diversity or combinations of objects encountered in the wild, and our findings may apply to real-world scenarios only to a certain extent. Future work should include more diverse participants and broaden the scope of objects to capture the complexity of everyday contexts better.

Methodologically, we relied mainly on synthetically generated images to eliminate possible external effects on perceived privacy. While generative AI is useful in HCI research and we instructed participants to imagine themselves as the subject or photographer, artificial images may not evoke the same cognitive responses as real ones. On the other hand, we studied privacy perception with at most two objects, whereas real-world visual content contains higher-order combinations; examining these could yield more detailed observations about co-presence and dominance effects. Furthermore, conditional branching reduced survey fatigue but resulted in too few responses in some branches, limiting statistical comparisons. Future work may complement synthetic stimuli with ethically curated real-world images and increase the sample size, ensuring sufficient coverage across all branches.




\section{Conclusion}

This paper advances the study of visual privacy by moving beyond a whole-image approach to a fine-grained, object-level perspective. Using a mixed-methods study with 92 participants, we uncovered how specific objects, their co-presence, and contextual cues shape people's privacy perceptions. Our results highlight the hidden dynamics that guide privacy judgments and demonstrate that small contextual details can substantially influence perceived privacy concerns.
We reemphasize the importance of human-centric design approaches that can help simplify the cognitive and technical complexity of visual privacy. Towards this end, we believe that our work lays the groundwork for developing adaptive, context-aware systems by providing empirical and practical insights. 


\bibliographystyle{IEEEtran}
%
\bibliography{sections/references}

@String{Computing = "Computing" }

@String{Computer = "{IEEE} Computer" }

@String{Springer = "Springer-Verlag" }

@article{10.1145/3517384,
author = {Stangl, Abigale and Shiroma, Kristina and Davis, Nathan and Xie, Bo and Fleischmann, Kenneth R. and Findlater, Leah and Gurari, Danna},
title = {Privacy Concerns for Visual Assistance Technologies},
year = {2022},
issue_date = {June 2022},
publisher = {Association for Computing Machinery},
address = {New York, NY, USA},
volume = {15},
number = {2},
issn = {1936-7228},
url = {https://doi.org/10.1145/3517384},
doi = {10.1145/3517384},
journal = {ACM Trans. Access. Comput.},
month = may,
articleno = {15},
numpages = {43}
}

@misc{photutorial2023,
  author       = {Matic Broz},
  title        = {How Many Photos Are Taken Every Day? (Updated 2023)},
  year         = {2023},
  url          = {https://photutorial.com/photos-statistics/},
  note         = {Accessed: 2025-06-04}
}

@article{burnes2020risk,
title = {Risk and Protective Factors of Identity Theft Victimization in the United States},
journal = {Preventive Medicine Reports},
volume = {17},
pages = {101058},
year = {2020},
issn = {2211-3355},
doi = {https://doi.org/10.1016/j.pmedr.2020.101058},
author = {David Burnes and Marguerite DeLiema and Lynn Langton},

}

@inproceedings{orekondy2017towards,
  title = {Towards a Visual Privacy Advisor: Understanding and Predicting Privacy Risks in Images},
  author = {Tribhuvanesh Orekondy and Bernt Schiele and Mario Fritz},
  year = {2017},
  date = {2017-10-29},
  booktitle = {IEEE International Conference on Computer Vision (ICCV)},
  pubstate = {published},
  tppubtype = {inproceedings}
}

@inproceedings{orekondy2018connecting,
    author = {Orekondy, Tribhuvanesh and Fritz, Mario and Schiele, Bernt},
    title = {Connecting Pixels to Privacy and Utility: Automatic Redaction of Private Information in Images},
    booktitle = {Conference on Computer Vision and Pattern Recognition (CVPR)},
    year = {2018}
}

@inproceedings{akter2020uncomfortable,
author = {Taslima Akter and Bryan Dosono and Tousif Ahmed and Apu Kapadia and Bryan Semaan},
title = {"I am Uncomfortable Sharing what I can{\textquoteright}t see": Privacy Concerns of the Visually Impaired with Camera Based Assistive Applications},
booktitle = {29th USENIX Security Symposium (USENIX Security 20)},
year = {2020},
isbn = {978-1-939133-17-5},
pages = {1929--1948},
url = {https://www.usenix.org/conference/usenixsecurity20/presentation/akter},
publisher = {USENIX Association},
month = aug
}

@techreport{amil2024impact,
  title={The Impact of AI-Driven Personalization Tools on Privacy Concerns and Consumer Trust in E-commerce},
  author={Amil, Yasmine and others},
  year={2024},
  source = {https://matheo.uliege.be/handle/2268.2/21371},
  publisher={Universit{\'e} de Li{\`e}ge, Li{\`e}ge, Belgique}
}

@article{nissenbaum2004privacy,
  title={Privacy as Contextual Integrity},
  author={Nissenbaum, Helen},
  journal={Wash. L. Rev.},
  volume={79},
  pages={119},
  year={2004},
  publisher={HeinOnline}
}

@article{hoyle2020privacynorms,
author = {Hoyle, Roberto and Stark, Luke and Ismail, Qatrunnada and Crandall, David and Kapadia, Apu and Anthony, Denise},
title = {Privacy Norms and Preferences for Photos Posted Online},
year = {2020},
issue_date = {August 2020},
publisher = {Association for Computing Machinery},
address = {New York, NY, USA},
volume = {27},
number = {4},
issn = {1073-0516},
url = {https://doi.org/10.1145/3380960},
journal = {ACM Trans. Comput.-Hum. Interact.},
month = aug,
articleno = {30},
numpages = {27},
}

@article{Chan03042019,
author = {Tommy K. H. Chan and Christy M. K. Cheung and Randy Y. M. Wong},
title = {Cyberbullying on Social Networking Sites: The Crime Opportunity and Affordance Perspectives},
journal = {Journal of Management Information Systems},
volume = {36},
number = {2},
pages = {574--609},
year = {2019},
publisher = {Routledge},
url = {     
        https://doi.org/10.1080/07421222.2019.1599500
}
}

@article{10.1145/3708501,
author = {Zhao, Ruoyu and Zhang, Yushu and Wang, Tao and Wen, Wenying and Xiang, Yong and Cao, Xiaochun},
title = {Visual Content Privacy Protection: A Survey},
year = {2025},
issue_date = {May 2025},
publisher = {Association for Computing Machinery},
address = {New York, NY, USA},
volume = {57},
number = {5},
issn = {0360-0300},
url = {https://doi.org/10.1145/3708501},
journal = {ACM Comput. Surv.},
month = jan,
articleno = {122},
numpages = {36},
}

@article{10.1145/3547299,
author = {Liu, Chi and Zhu, Tianqing and Zhang, Jun and Zhou, Wanlei},
title = {Privacy Intelligence: A Survey on Image Privacy in Online Social Networks},
year = {2022},
issue_date = {August 2023},
publisher = {Association for Computing Machinery},
address = {New York, NY, USA},
volume = {55},
number = {8},
issn = {0360-0300},
url = {https://doi.org/10.1145/3547299},
journal = {ACM Comput. Surv.},
month = dec,
articleno = {161},
numpages = {35}
}

@article{goyeneche2024linked,
    author = {Goyeneche, David and Singaraju, Stephen and Arango, Luis},
    title = {Linked by Age: A Study on Social Media Privacy Concerns Among Younger and Older Adults},
    journal = {Industrial Management \& Data Systems},
    volume = {124},
    number = {2},
    pages = {640-665},
    year = {2023},
    month = {12},
    issn = {0263-5577},
    url = {https://doi.org/10.1108/IMDS-07-2023-0462}
}

@inproceedings{10.1145/2396761.2398735,
author = {Zerr, Sergej and Siersdorfer, Stefan and Hare, Jonathon},
title = {PicAlert! A System for Privacy-Aware Image Classification and Retrieval},
year = {2012},
isbn = {9781450311564},
publisher = {Association for Computing Machinery},
address = {New York, NY, USA},
url = {https://doi.org/10.1145/2396761.2398735},
booktitle = {Proceedings of the 21st ACM International Conference on Information and Knowledge Management},
pages = {2710–2712},
numpages = {3},
location = {Maui, Hawaii, USA},
series = {CIKM '12}
}

@INPROCEEDINGS{Gurari2019VizWiz,
  author={Gurari, Danna and Li, Qing and Lin, Chi and Zhao, Yinan and Guo, Anhong and Stangl, Abigale and Bigham, Jeffrey P.},
  booktitle={2019 IEEE/CVF Conference on Computer Vision and Pattern Recognition (CVPR)}, 
  title={VizWiz-Priv: A Dataset for Recognizing the Presence and Purpose of Private Visual Information in Images Taken by Blind People}, 
  year={2019},
  volume={},
  number={},
  pages={939-948},
  doi={10.1109/CVPR.2019.00103}}

@inproceedings{Niu2025Bystanders,
author = {Niu, Yuqi and Meng-Schneider, Nicole and Qiu, Weidong and Kokciyan, Nadin},
title = {``I am not the primary focus" - Understanding the Perspectives of Bystanders in Photos Shared Online},
year = {2025},
isbn = {9798400713941},
publisher = {Association for Computing Machinery},
address = {New York, NY, USA},
url = {https://doi.org/10.1145/3706598.3713826},
articleno = {899},
numpages = {23},
location = {
},
series = {CHI '25}
}

@techreport{redmiles2017summary,
  title={A Summary of Survey Methodology Best Practices for Security and Privacy Researchers},
  author={Redmiles, Elissa M and Acar, Yasemin and Fahl, Sascha and Mazurek, Michelle L},
  year={2017},
institution={University of Maryland}
}

@inproceedings{egelman2007security,
author = {Egelman, Serge and King, Jennifer and Miller, Robert C. and Ragouzis, Nick and Shehan, Erika},
title = {Security User Studies: Methodologies and Best Practices},
year = {2007},
isbn = {9781595936424},
publisher = {Association for Computing Machinery},
address = {New York, NY, USA},
url = {https://doi.org/10.1145/1240866.1241089},
booktitle = {CHI '07 Extended Abstracts on Human Factors in Computing Systems},
pages = {2833–2836},
numpages = {4},
location = {San Jose, CA, USA},
series = {CHI EA '07}
}

@book{rea2014designing,
  title={Designing and Conducting Survey Research: A Comprehensive Guide},
  author={Rea, Louis M and Parker, Richard A},
  year={2014},
  publisher={John Wiley \& Sons}
}

@article{Tonge2020Image,
author = {Tonge, Ashwini and Caragea, Cornelia},
title = {Image Privacy Prediction Using Deep Neural Networks},
year = {2020},
issue_date = {May 2020},
publisher = {Association for Computing Machinery},
address = {New York, NY, USA},
volume = {14},
number = {2},
issn = {1559-1131},
url = {https://doi.org/10.1145/3386082},
journal = {ACM Trans. Web},
month = apr,
articleno = {7},
numpages = {32},
}

@InProceedings{patwari2024Perceptanon,
  title = 	 {{P}ercept{A}non: Exploring the Human Perception of Image Anonymization Beyond Pseudonymization for {GDPR}},
  author =       {Patwari, Kartik and Chuah, Chen-Nee and Lyu, Lingjuan and Sharma, Vivek},
  booktitle = 	 {Proceedings of the 41st International Conference on Machine Learning},
  pages = 	 {39955--39971},
  year = 	 {2024},
  editor = 	 {Salakhutdinov, Ruslan and Kolter, Zico and Heller, Katherine and Weller, Adrian and Oliver, Nuria and Scarlett, Jonathan and Berkenkamp, Felix},
  volume = 	 {235},
  series = 	 {Proceedings of Machine Learning Research},
  month = 	 {21--27 Jul},
  publisher =    {PMLR},
}

@incollection{braun2023doing,
  title={Doing Reflexive Thematic Analysis},
  author={Braun, Virginia and Clarke, Victoria and Hayfield, Nikki and Davey, Louise and Jenkinson, Elizabeth},
  booktitle={Supporting Research in Counselling and Psychotherapy: Qualitative, Quantitative, and Mixed Methods Research},
  pages={19--38},
  year={2023},
  publisher={Springer}
}

@book{blandford2016qualitative,
  title={Qualitative HCI Research: Going Behind the Scenes},
  author={Blandford, Ann and Furniss, Dominic and Makri, Stephann},
  year={2016},
  publisher={Morgan \& Claypool Publishers}
}

@article{mcdonald2019reliability,
author = {McDonald, Nora and Schoenebeck, Sarita and Forte, Andrea},
title = {Reliability and Inter-rater Reliability in Qualitative Research: Norms and Guidelines for CSCW and HCI Practice},
year = {2019},
issue_date = {November 2019},
publisher = {Association for Computing Machinery},
address = {New York, NY, USA},
volume = {3},
number = {CSCW},
url = {https://doi.org/10.1145/3359174},
month = nov,
articleno = {72},
numpages = {23},
journal = {Proc. ACM Hum.-Comput. Interact.},
}

@article{Belanger2016Privacy,
 ISSN = {02767783},
 URL = {http://www.jstor.org/stable/41409971},
 author = {France Bélanger and Robert E. Crossler},
 journal = {MIS Quarterly},
 number = {4},
 pages = {1017--1041},
 publisher = {Management Information Systems Research Center, University of Minnesota},
 title = {Privacy in the Digital Age: A Review of Information Privacy Research in Information Systems},
 urldate = {2025-09-10},
 volume = {35},
 year = {2011}
}

@article{kaur2021systematic,
  title={A Systematic Literature Review on Cyberstalking. An Analysis of Past Achievements and Future Promises},
  author={Kaur, Puneet and Dhir, Amandeep and Tandon, Anushree and Alzeiby, Ebtesam A and Abohassan, Abeer Ahmed},
  journal={Technological Forecasting and Social Change},
  volume={163},
  pages={120426},
  year={2021},
  publisher={Elsevier}
}

@misc{ww2photomystery2023,
  author       = {Mossou, Annique},
  title        = {Solving World War II Photo Mysteries With Open Source Techniques},
  year         = {2023},
  url          = {https://www.bellingcat.com/news/2023/08/04/solving-world-war-ii-photo-mysteries-with-open-source-techniques/},
  note         = {Accessed: 2025-09-09}
}

@misc{chronolocation2023,
  author       = {Weide, Youri van der},
  title        = {Chronolocation: Determining When a Photo was Taken Using Facebook, Google Street View and Assorted Tiny Details },
  year         = {2023},
  url          = {https://www.bellingcat.com/resources/2023/05/08/chronolocation-determining-when-a-photo-was-taken-using-facebook-google-street-view-and-assorted-tiny-details/},
  note         = {Accessed: 2025-09-09}
}

@misc{ChinaShips2020,
  author       = {Urbina, Ian},
  title        = {The Deadly Secret of China's Invisible Armada},
  year         = {2020},
  url          = {https://www.nbcnews.com/specials/china-illegal-fishing-fleet/},
  note         = {Accessed: 2025-09-09}
}

@article{ravi2024review,
  title={A Review on Visual Privacy Preservation Techniques for Active and Assisted living},
  author={Ravi, Siddharth and Climent-P{\'e}rez, Pau and Florez-Revuelta, Francisco},
  journal={Multimedia Tools and Applications},
  volume={83},
  number={5},
  pages={14715--14755},
  year={2024},
  publisher={Springer},
  url={https://doi.org/10.1007/s11042-023-15775-2}
}

@inproceedings{stangl2020visual,
author = {Stangl, Abigale and Shiroma, Kristina and Xie, Bo and Fleischmann, Kenneth R. and Gurari, Danna},
title = {Visual Content Considered Private by People Who are Blind},
year = {2020},
isbn = {9781450371032},
publisher = {Association for Computing Machinery},
address = {New York, NY, USA},
url = {https://doi.org/10.1145/3373625.3417014},
booktitle = {Proceedings of the 22nd International ACM SIGACCESS Conference on Computers and Accessibility},
articleno = {31},
numpages = {12},
location = {Virtual Event, Greece},
series = {ASSETS '20}
}

@inproceedings{zhang2024designing,
author = {Zhang, Lotus and Stangl, Abigale and Sharma, Tanusree and Tseng, Yu-Yun and Xu, Inan and Gurari, Danna and Wang, Yang and Findlater, Leah},
title = {Designing Accessible Obfuscation Support for Blind Individuals’ Visual Privacy Management},
year = {2024},
isbn = {9798400703300},
publisher = {Association for Computing Machinery},
address = {New York, NY, USA},
url = {https://doi.org/10.1145/3613904.3642713},
booktitle = {Proceedings of the 2024 CHI Conference on Human Factors in Computing Systems},
articleno = {235},
numpages = {19},
location = {Honolulu, HI, USA},
series = {CHI '24}
}

@inproceedings{akter2020privacy,
author = {Akter, Taslima and Ahmed, Tousif and Kapadia, Apu and Swaminathan, Swami Manohar},
title = {Privacy Considerations of the Visually Impaired with Camera Based Assistive Technologies: Misrepresentation, Impropriety, and Fairness},
year = {2020},
isbn = {9781450371032},
publisher = {Association for Computing Machinery},
address = {New York, NY, USA},
url = {https://doi.org/10.1145/3373625.3417003},
booktitle = {Proceedings of the 22nd International ACM SIGACCESS Conference on Computers and Accessibility},
articleno = {32},
numpages = {14},
location = {Virtual Event, Greece},
series = {ASSETS '20}
}

@incollection{wisniewski2022privacy,
  title={Privacy theories and frameworks},
  author={Wisniewski, Pamela J and Page, Xinru},
  booktitle={Modern socio-technical perspectives on privacy},
  pages={15--41},
  year={2022},
  publisher={Springer International Publishing Cham}
}

@book{petronio2002boundaries,
  title={Boundaries of privacy: Dialectics of disclosure},
  author={Petronio, Sandra},
  year={2002},
  publisher={Suny Press}
}

@incollection{acquisti2007can,
  title={What can behavioral economics teach us about privacy?},
  author={Acquisti, Alessandro and Grossklags, Jens},
  booktitle={Digital privacy},
  pages={363--378},
  year={2007},
  publisher={Auerbach Publications}
}

@article{dinev2006extended,
  title={An extended privacy calculus model for e-commerce transactions},
  author={Dinev, Tamara and Hart, Paul},
  journal={Information systems research},
  volume={17},
  number={1},
  pages={61--80},
  year={2006},
  publisher={Informs}
}

@inproceedings{kairam2016snap,
  title={Snap decisions? How users, content, and aesthetics interact to shape photo sharing behaviors},
  author={Kairam, Sanjay and Kaye, Joseph'Jofish' and Guerra-Gomez, John Alexis and Shamma, David A},
  booktitle={Proceedings of the 2016 CHI Conference on Human Factors in Computing Systems},
  pages={113--124},
  year={2016}
}

@inproceedings{ahern2007over,
  title={Over-exposed? Privacy patterns and considerations in online and mobile photo sharing},
  author={Ahern, Shane and Eckles, Dean and Good, Nathaniel S and King, Simon and Naaman, Mor and Nair, Rahul},
  booktitle={Proceedings of the SIGCHI conference on Human factors in computing systems},
  pages={357--366},
  year={2007}
}

@inproceedings{habib2019impact,
  title={Impact of contextual factors on snapchat public sharing},
  author={Habib, Hana and Shah, Neil and Vaish, Rajan},
  booktitle={Proceedings of the 2019 CHI Conference on Human Factors in Computing Systems},
  pages={1--13},
  year={2019}
}

@inproceedings{li2020towards,
  title={Towards a taxonomy of content sensitivity and sharing preferences for photos},
  author={Li, Yifang and Vishwamitra, Nishant and Hu, Hongxin and Caine, Kelly},
  booktitle={Proceedings of the 2020 CHI Conference on Human Factors in Computing Systems},
  pages={1--14},
  year={2020}
}

@inproceedings{vishwamitra2017blur,
  title={Blur vs. block: Investigating the effectiveness of privacy-enhancing obfuscation for images},
  author={Vishwamitra, Nishant and Knijnenburg, Bart and Hu, Hongxin and Kelly Caine, Yifang P and others},
  booktitle={Proceedings of the IEEE Conference on Computer Vision and Pattern Recognition Workshops},
  pages={39--47},
  year={2017}
}

@inproceedings{khamis2022deepfakes,
  title={DeepFakes for privacy: Investigating the effectiveness of state-of-the-art privacy-enhancing face obfuscation methods},
  author={Khamis, Mohamed and Farzand, Habiba and Mumm, Marija and Marky, Karola},
  booktitle={Proceedings of the 2022 International Conference on Advanced Visual Interfaces},
  pages={1--5},
  year={2022}
}

@inproceedings{khamis2024perspectives,
  title={Perspectives on DeepFakes for Privacy: Comparing Perceptions of Photo Owners and Obfuscated Individuals towards DeepFake Versus Traditional Privacy-Enhancing Obfuscation},
  author={Khamis, Mohamed and Panskus, Rebecca and Farzand, Habiba and Mumm, Marija and Macdonald, Shaun and Marky, Karola},
  booktitle={Proceedings of the International Conference on Mobile and Ubiquitous Multimedia},
  pages={300--312},
  year={2024}
}

@article{li2017effectiveness,
  title={Effectiveness and users' experience of obfuscation as a privacy-enhancing technology for sharing photos},
  author={Li, Yifang and Vishwamitra, Nishant and Knijnenburg, Bart P and Hu, Hongxin and Caine, Kelly},
  journal={Proceedings of the ACM on Human-Computer Interaction},
  volume={1},
  number={CSCW},
  pages={1--24},
  year={2017},
  publisher={ACM New York, NY, USA}
}

@article{wen2024image,
  title={Image Privacy Protection: A Survey},
  author={Wen, Wenying and Yuan, Ziye and Zhang, Yushu and Wang, Tao and Xiao, Xiangli and Zhao, Ruoyu and Fang, Yuming},
  journal={arXiv preprint arXiv:2412.15228},
  year={2024}
}

@article{lu2023seeing,
  title={Seeing is not always believing: Benchmarking human and model perception of ai-generated images},
  author={Lu, Zeyu and Huang, Di and Bai, Lei and Qu, Jingjing and Wu, Chengyue and Liu, Xihui and Ouyang, Wanli},
  journal={Advances in neural information processing systems},
  volume={36},
  pages={25435--25447},
  year={2023}
}

@article{lago2021more,
  title={More real than real: A study on human visual perception of synthetic faces [applications corner]},
  author={Lago, Federica and Pasquini, Cecilia and B{\"o}hme, Rainer and Dumont, H{\'e}l{\`e}ne and Goffaux, Val{\'e}rie and Boato, Giulia},
  journal={IEEE Signal Processing Magazine},
  volume={39},
  number={1},
  pages={109--116},
  year={2021},
  publisher={IEEE}
}

@article{nightingale2021synthetic,
  title={Synthetic faces: how perceptually convincing are they?},
  author={Nightingale, Sophie and Agarwal, Shruti and H{\"a}rk{\"o}nen, Erik and Lehtinen, Jaakko and Farid, Hany},
  journal={Journal of vision},
  volume={21},
  number={9},
  pages={2015--2015},
  year={2021},
  publisher={The Association for Research in Vision and Ophthalmology}
}

@inproceedings{shen2021study,
  title={A study of the human perception of synthetic faces},
  author={Shen, Bingyu and RichardWebster, Brandon and O'Toole, Alice and Bowyer, Kevin and Scheirer, Walter J},
  booktitle={2021 16th IEEE International Conference on Automatic Face and Gesture Recognition (FG 2021)},
  pages={1--8},
  year={2021},
  organization={IEEE}
}

@inproceedings{bilucaglia2024emotional,
  title={Emotional reactions to AI-generated images: a pilot study using neurophysiological measures},
  author={Bilucaglia, Marco and Casiraghi, Chiara and Bruno, Alessandro and Chiarelli, Simone and Fici, Alessandro and Russo, Vincenzo and Zito, Margherita},
  booktitle={International Conference on Machine Learning, Optimization, and Data Science},
  pages={147--161},
  year={2024},
  organization={Springer}
}

@article{boerman2017online,
  title={Online behavioral advertising: A literature review and research agenda},
  author={Boerman, Sophie C and Kruikemeier, Sanne and Zuiderveen Borgesius, Frederik J},
  journal={Journal of advertising},
  volume={46},
  number={3},
  pages={363--376},
  year={2017},
  publisher={Taylor \& Francis}
}

@article{puglisi2017web,
  title={On web user tracking of browsing patterns for personalised advertising},
  author={Puglisi, Silvia and Rebollo-Monedero, David and Forn{\'e}, Jordi},
  journal={International Journal of Parallel, Emergent and Distributed Systems},
  volume={32},
  number={5},
  pages={502--521},
  year={2017},
  publisher={Taylor \& Francis}
}

@article{malhotra2004internet,
  title={Internet users' information privacy concerns (IUIPC): The construct, the scale, and a causal model},
  author={Malhotra, Naresh K and Kim, Sung S and Agarwal, James},
  journal={Information systems research},
  volume={15},
  number={4},
  pages={336--355},
  year={2004},
  publisher={Informs}
}

@article{cheung2016evaluating,
  title={Evaluating the privacy risk of user-shared images},
  author={Cheung, Ming and She, James},
  journal={ACM Transactions on Multimedia Computing, Communications, and Applications (TOMM)},
  volume={12},
  number={4s},
  pages={1--21},
  year={2016},
  publisher={ACM New York, NY, USA}
}

@article{barnes2006privacy,
  title={A privacy paradox: Social networking in the United States},
  author={Barnes, Susan B},
  journal={First Monday},
  year={2006}
}

@article{gerber2018explaining,
  title={Explaining the privacy paradox: A systematic review of literature investigating privacy attitude and behavior},
  author={Gerber, Nina and Gerber, Paul and Volkamer, Melanie},
  journal={Computers \& security},
  volume={77},
  pages={226--261},
  year={2018},
  publisher={Elsevier}
}

@article{chiang2025understanding,
  title={Understanding User Needs and Attitudes for Privacy Protection Tools in Online Visual Content Sharing},
  author={Chiang, Chun-Wei and Tian, Harry Yizhou and Yin, Ming},
  journal={Proceedings of the ACM on Human-Computer Interaction},
  volume={9},
  number={7},
  pages={1--31},
  year={2025},
  publisher={ACM New York, NY, USA}
}

\appendix


\section{Appendix}

\subsection{Prompts Used for Image Generation}
\label{app:prompts}

The text prompts used to generate the example images shown in 
Figure~\ref{fig:cafe_office_examples}. For each environment, the first image was generated from a base prompt, while subsequent variations were produced using SORA’s remix functionality.

\begin{itemize}[leftmargin=*]
    \item Figure 2a (Base image for café, face foreground only): \textit{
``A realistic indoor café scene in the late morning, softly illuminated by natural daylight streaming through large street-facing windows. A woman in her late 20s is seated alone at a small round table near the window, angled slightly toward the light. She has shoulder-length dark auburn hair, loosely tied back, and wears a cream-colored short-sleeve knit top and dark jeans. Her forearms rest naturally on the table, making them clearly visible in the frame. Her skin is light olive-toned, and she has a calm, neutral expression as she gazes out the window with a subtle half-smile, unaware of the camera. On the table are a few personal items: a half-full ceramic mug, a closed leather-bound notebook, and a smartphone lying screen-down. The table itself is a worn, wooden surface with a bit of character—subtle scratches and warm tones. She sits on a simple wooden chair with a low back, and there’s a canvas tote bag hanging off the side. In the softly blurred background, the café reveals other details: a brick accent wall, a large chalkboard menu partially visible above the counter, and some framed posters and event flyers loosely pinned to a corkboard near the entrance. There are a few other patrons seated further back—some chatting, some working on laptops—but none are clearly distinguishable. The camera angle is natural and intimate, positioned at eye level and slightly off-center, capturing the woman from the front-left in a three-quarter view. Her arms are fully visible on the tabletop. The lighting is warm, realistic, and casts soft, diffused shadows across the scene. The photograph feels candid and everyday—an ordinary moment caught in passing, creating a grounded and relatable atmosphere.''}  
    \item Figure 2b (Face + medicine): \textit{On the table there is medicine.}  
    \item Figure 2c (Face + medicine + landmark): \textit{On the wall in the back, the cafe name is written: ``Elm Street Cafe.''}  
    \item Figure 2d (Base image for office, face foreground only): \textit{
``A photorealistic indoor office scene set in a mid-size, modern corporate workspace during a weekday morning. The environment is structured and clean, with carpeted floors, partitioned desks, and frosted glass panels along the corridor wall. Overhead lighting casts a neutral white tone across the space, supplemented by soft daylight from windows with mesh roller blinds half-drawn. In the foreground, a man in his early 30s is seated alone at an L-shaped desk inside a semi-open cubicle. He has short dark brown hair, light stubble, and wears a dark gray button-up shirt with a lanyard ID badge around his neck. His posture is engaged but relaxed, and he’s focused on his dual-monitor workstation — one screen shows a spreadsheet, the other a messaging app. His desk has typical office clutter: a keyboard, notepad with scribbled notes, ceramic coffee mug, a phone dock, and a small branded desk calendar. A jacket is draped over the back of his ergonomic chair, and there’s a cable tray visible beneath the desk. The background includes blurred silhouettes of other cubicles, vertical storage cabinets, and a meeting room with glass doors partially open. The tone is realistic and corporate — capturing a candid moment of one employee at work, with the environment grounded in everyday office detail.''} 
    \item Figure 2e (Face + tattoo): \textit{Add a tiger tattoo to the forearm, ``legend'' should be written below tiger.}  
    \item Figure 2f (Face + tattoo + face (a.)): \textit{Add a person.}  
\end{itemize}

\subsection{Full Survey (Café Version)}
\label{appendix:full_survey_cafe}

\noindent{Note: Office version of the survey has the same questions. The only difference is that the displayed AI-generated images show a working office environment.}
\vspace{0.8em}
\noindent\textbf{Page 0: Instructions} 

\noindent{Participants were presented with a brief overview of the survey's context and informed that all images were synthetic and generated using AI tools, with no real individuals or personal data involved. Information on data processing was disclosed, including the names of the responsible parties, the types of personal data collected (e.g., demographic data and survey responses) and the applicable legal basis under GDPR. Participants were informed of their rights regarding access, correction, and deletion of their data, and withdrawal of consent, along with instructions for submitting privacy-related requests or complaints. To start the survey, participants were required to give their informed consent and enter their email address for validation purposes. Optionally, they could choose to enter the lottery to win a movie ticket.}

\vspace{0.8em}
\noindent\textbf{Page 1: Demographic Information} 

\noindent{Please answer a few demographic questions.}

\vspace{1em}
\noindent\textbf{Q1. What is your age?} \\
18--24 \\
25--34 \\
35--44 \\
45--54 \\
55 and older

\vspace{0.8em}
\noindent\textbf{Q2. How often do you share photos online?} \\
Never \\
Rarely \\
Sometimes \\
Often \\
Daily

\vspace{0.8em}
\noindent\textbf{Q3. What type of devices do you regularly use? (Select all that apply)} \\
Mobile phone \\
Laptop/Desktop computer \\
Smart watch \\
Fitness tracker \\
Wearable devices (e.g., heart rate monitor, VR) \\
Other

\vspace{0.8em}
\noindent\textbf{Q4. What is your ethnic background? (Select one)} \\
White / Caucasian \\
Black or African \\
Asian \\
Native American / Indigenous \\
Mixed / Multi \\
I don't wish to disclose \\
Other

\vspace{0.8em}
\noindent\textbf{Q5. What is your gender?} \\
Male \\
Female \\
Other \\
Prefer not to say

\vspace{0.8em}
\noindent\textbf{Q6. What is your highest level of education?} \\
Less than high school \\
High school graduate \\
Some college \\
2-year degree \\
Bachelor's degree \\
Master's degree \\
Doctorate

\vspace{0.8em}
\noindent\textbf{Q7. What is your professional background?} \\
Employed full-time \\
Employed part-time \\
Unemployed (seeking) \\
Unemployed (not seeking) \\
Retired \\
Student

\vspace{0.8em}
\noindent\textbf{Q8-11. Privacy Preference Statements} \\
Please indicate your agreement with the following statements:

\begin{tabular}{@{}p{10.5cm}l@{}}
(a) I am concerned about my privacy online & [1–5] Likert scale \\
(b) I am concerned about my privacy in everyday life & [1–5] Likert scale \\
(c) It bothers me to give personal info to many online companies & [1–5] Likert scale \\
(d) I check photos for personal info before sharing online & [1–5] Likert scale \\
\end{tabular}

\vspace{0.5em}
\noindent\textit{Scale: 1 = Disagree, 2 = Somewhat Disagree, 3 = Neutral, 4 = Somewhat Agree, 5 = Agree}

\vspace{0.8em}

\noindent\textbf{Page 2: Visual Privacy – Face} 

\noindent{Suppose you are sitting at a table in a cafeteria. For each of the below groups of images, picture yourself as the subject in the photograph. Either you are the one taking the photo from your perspective, or someone takes a picture of you.} 

\vspace{0.8em}
\noindent\textbf{Q12. How comfortable would you feel sharing this photo of yourself on your social media account?} 

\begin{figure}[t]
\centering
\includegraphics[width=0.8\columnwidth]{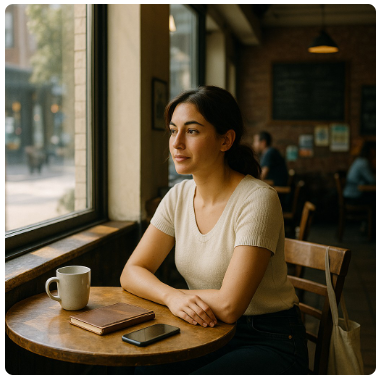}
\caption{Visual accompanying Q12. \textit{Scale: 1 = Not comfortable at all, 2 = Slightly uncomfortable, 3 = Neutral, 4 = Slightly comfortable, 5 = Very comfortable}}



\end{figure}

\vspace{0.8em}
\noindent\textbf{Q13-17. Suppose while taking the picture, there were some other objects captured along with the background. How comfortable would you feel sharing the picture of yourself if the following objects were in the image?} 

\begin{figure}[t]
\begin{subfigure}{0.45\columnwidth}
    \includegraphics[width=\textwidth, trim={0 0 19cm 0},clip]{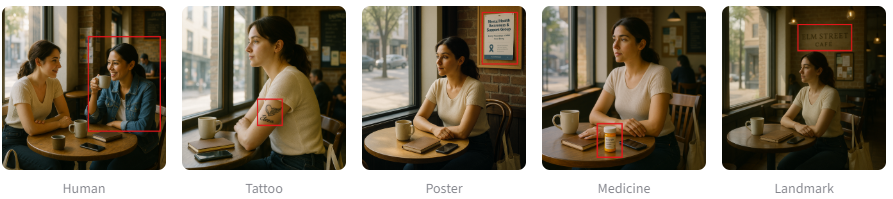}
    \caption{[1–5] Likert scale}
\end{subfigure}
\hfill
\begin{subfigure}{0.45\columnwidth}
    \includegraphics[width=\textwidth, trim={4.75cm 0 14.25cm 0},clip]{images/person22.png}
    \caption{[1–5] Likert scale} 
\end{subfigure} 
\begin{subfigure}{0.45\columnwidth} 
    \includegraphics[width=\textwidth, trim={9.5cm 0 9.5cm 0},clip]{images/person22.png} 
    \caption{[1–5] Likert scale} 
\end{subfigure}  
\hfill 
\begin{subfigure}{0.45\columnwidth} 
    \includegraphics[width=\textwidth, trim={14.25cm 0 4.75cm 0},clip]{images/person22.png} 
    \caption{[1–5] Likert scale} 
\end{subfigure} 
\hfill 
    \begin{center}
\begin{subfigure}{0.45\columnwidth} 
    \includegraphics[width=\textwidth, trim={19cm 0 0 0},clip]{images/person22.png} 
    \caption{[1–5] Likert scale} 
\end{subfigure} 
    \end{center}
    \caption{Visuals accompanying Q13-17. \textit{Likert Scale: 1 = Not comfortable at all, 2 = Slightly uncomfortable, 3 = Neutral, 4 = Slightly comfortable, 5 = Very comfortable}}
\end{figure}



%



\vspace{0.8em}
\noindent\textbf{Q18. Based on your answers above, please rank the objects in order of sensitivity.} \\
Human \quad Tattoo \quad Poster \quad Medicine \quad Landmark

\vspace{0.5em}
\noindent\textit{1 = Most sensitive, 5 = Least Sensitive}

\vspace{0.8em}
\noindent\textbf{Q19. Why did you rank the objects in this order?} (Open text)

\vspace{0.8em}
\noindent\textbf{Q20-23. (Assume that the participant ranked the landmark as the background PSO with the least effect on the comfortability level) You realized that the photo including the landmark also included the following objects. How comfortable would you feel sharing the picture of yourself that also contains landmark if the following objects were in the image as well?} 

\begin{figure}[t]
\begin{subfigure}{0.45\columnwidth}
    \includegraphics[width=\textwidth, trim={0 0 18cm 0},clip]{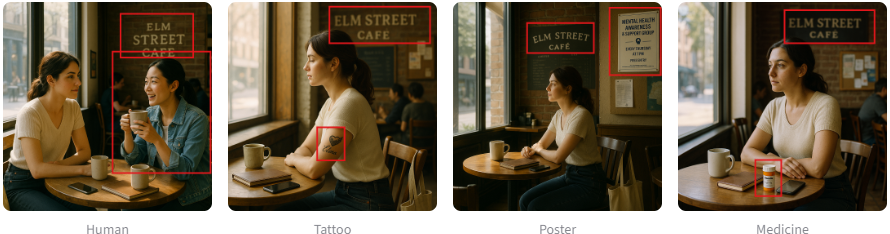}
    \caption{[1–5] Likert scale}
\end{subfigure}
\hfill
\begin{subfigure}{0.45\columnwidth}
    \includegraphics[width=\textwidth, trim={6cm 0 12cm 0},clip]{images/person33.png}
    \caption{[1–5] Likert scale} 
\end{subfigure} 
\begin{subfigure}{0.45\columnwidth} 
    \includegraphics[width=\textwidth, trim={12cm 0 6cm 0},clip]{images/person33.png} 
    \caption{[1–5] Likert scale} 
\end{subfigure}  
\hfill 
\begin{subfigure}{0.45\columnwidth} 
    \includegraphics[width=\textwidth, trim={18cm 0 0cm 0},clip]{images/person33.png} 
    \caption{[1–5] Likert scale} 
\end{subfigure} 
\hfill 
    \caption{Visuals accompanying Q20-23. \textit{Likert Scale: 1 = Not comfortable at all, 2 = Slightly uncomfortable, 3 = Neutral, 4 = Slightly comfortable, 5 = Very comfortable}}
\end{figure}
 







\vspace{0.8em}
\noindent\textbf{Q24. Can you explain what the effect of landmark was on how comfortable you felt sharing the picture of yourself with the other objects?} (Open text) 
\vspace{0.8em}

\noindent\textbf{Page 3: Visual Privacy – Miscellaneous Paper}

\noindent{The structure of this page mirrors that of the previous section (face as a foreground PSO). Participants are asked to evaluate their comfortability levels with images involving \textbf{Miscellaneous Paper} as the foreground PSO. After rating their comfort with the foreground PSO alone, the participants are asked the same questions where the background PSOs are \textbf{face (photo), full name, birthdate, signature}, and \textbf{address}. As before, they complete individual Likert scale ratings, rank the background items according to their effect on the comfortability level, and rate the effect of the combination of background PSOs. This section includes questions \textbf{Q25–Q37}.}

\vspace{0.8em}

\noindent\textbf{Page 4: Visual Privacy – Screen} 

{The structure of this page mirrors that of the previous section (face as a foreground PSO). Participants are asked to evaluate their comfortability levels with images involving \textbf{Computer/Phone screen} as the foreground PSO. After rating their comfort with the foreground PSO alone, the participants are asked the same questions where the background PSOs are \textbf{face (reflection), email, username, date}, and \textbf{phone number}. As before, they complete individual Likert scale ratings, rank the background items according to their effect on the comfortability level, and rate the effect of the combination of background PSOs. This section includes questions \textbf{Q38–Q50}.}

\vspace{0.8em}

\noindent\textbf{Page 5: Review and Submit} 

\noindent{Participants were shown a summary of their responses and asked to confirm and submit.}

\subsection{Additional Tables and Figures}
\label{appendix:additional_tables}

\begin{table}[h]
\centering
\caption{Mapping between comfortability levels and perceived privacy levels.}
\label{tab:comfort_privacy_mapping}
\begin{adjustbox}{width=0.96\columnwidth}
\begin{tabular}{cc}
\toprule
\textbf{Comfortability Level} & \textbf{Perceived Privacy Level} \\
\midrule
1 (Not comfortable at all) & 5 (Most sensitive) \\
2 (Slightly uncomfortable) & 4 \\
3 (Neutral) & 3 \\
4 (Slightly comfortable) & 2 \\
5 (Very comfortable) & 1 (Least sensitive) \\
\bottomrule
\end{tabular}
\end{adjustbox}
\end{table}

\begin{table}[t]
\centering
\caption{Demographic distribution of survey participants (n=92).}
\label{tab:demographics}
\begin{adjustbox}{width=0.85\columnwidth}
\begin{tabular}{l l c}
\toprule
Category & Response & Count (\%) \\
\midrule
Ethnicity & White/Caucasian & 58 (63\%)  \\
          & Asian & 20 (22\%)           \\
          & Black/African & 6 (7\%)   \\
          & Mixed/Multi & 3 (3\%) \\
          & I don't wish to disclose & 5 (5\%) \\        
\midrule
Age & 18--24 & 18 (20\%)   \\
    & 25--34 & 32 (35\%)           \\
    & 35--44 & 9 (10\%)             \\
    & 45--54 & 29 (32\%)            \\
    & 55+    & 4 (4\%)             \\
\midrule
Gender & Female & 47 (51\%) \\
       & Male & 43 (47\%)   \\
       & Other & 2 (2\%) \\
\midrule
Education & High school graduate & 4 (4\%) \\
          & Some college & 3 (3\%) \\
          & Bachelor's degree & 32 (35\%) \\
          & Master's degree & 29 (32\%) \\
          & Doctorate & 24 (26\%) \\  
\midrule
Profession & Employed full-time & 52 (57\%) \\
    & Employed part-time & 8 (9\%) \\
    & Student            & 28 (30\%) \\
    & Retired            & 2 (2\%) \\
    & Unemployed         & 2 (2\%) \\
\bottomrule
\end{tabular}
\end{adjustbox}
\end{table}

\begin{table}[t]
\centering
\caption{Significant demographic effects on comfortability levels. Mann–Whitney U tests were applied for gender, showing that female participants reported higher comfortability levels in all significant cases. Kruskal–Wallis H tests were used for the remaining demographic variables, indicating at least one group distribution differed significantly from the others.}
\label{tab:sig_demographics}
\begin{adjustbox}{width=1.0\columnwidth}
\begin{tabular}{lllc}
\toprule
Foreground PSO & Background PSO & Category & $p$-value \\
\midrule
\multicolumn{4}{l}{\textbf{Café}} \\
\midrule
Screen & E-mail      & Gender (F $>$ M)       & 0.0489 \\
Screen & Phone Number      & Gender (F $>$ M)       & 0.0289 \\
Face   & None       & Social Usage & 0.0332 \\
Paper  & Face (p.)      & Social Usage  & 0.0126 \\
Screen & None       & Social Usage  & 0.0391 \\
Screen & Face (r.)  & Social Usage  & 0.0401 \\
Face   & Landmark   & Profession    & 0.0377 \\
Screen & None       & Profession    & 0.0280 \\
Screen & Face (r.)  & Profession   & 0.0430 \\
Face   & Landmark   & Age       & 0.0206 \\
Paper  & Full Name       & Age          & 0.0355 \\
Screen & E-mail      & Age          & 0.0177 \\
Screen & Face (r.)  & Age          & 0.0206 \\
\midrule
\multicolumn{4}{l}{\textbf{Office}} \\
\midrule
Face   & Face (a.)      & Ethnicity    & 0.0399 \\
Paper  & Full Name       & Ethnicity    & 0.0056 \\
Paper  & Signature  & Ethnicity    & 0.0391 \\
Screen & E-mail      & Ethnicity   & 0.0012 \\
Paper  & Face (p.)      & Education    & 0.0427 \\
Face   & Tattoo     & Age         & 0.0393 \\
Screen & Date       & Age         & 0.0408 \\
\bottomrule
\end{tabular}
\end{adjustbox}
\end{table}

%

\newpage
\begin{figure*}[t]
    \centering
    \vspace{-35em}
    \includegraphics[width=0.9\linewidth]{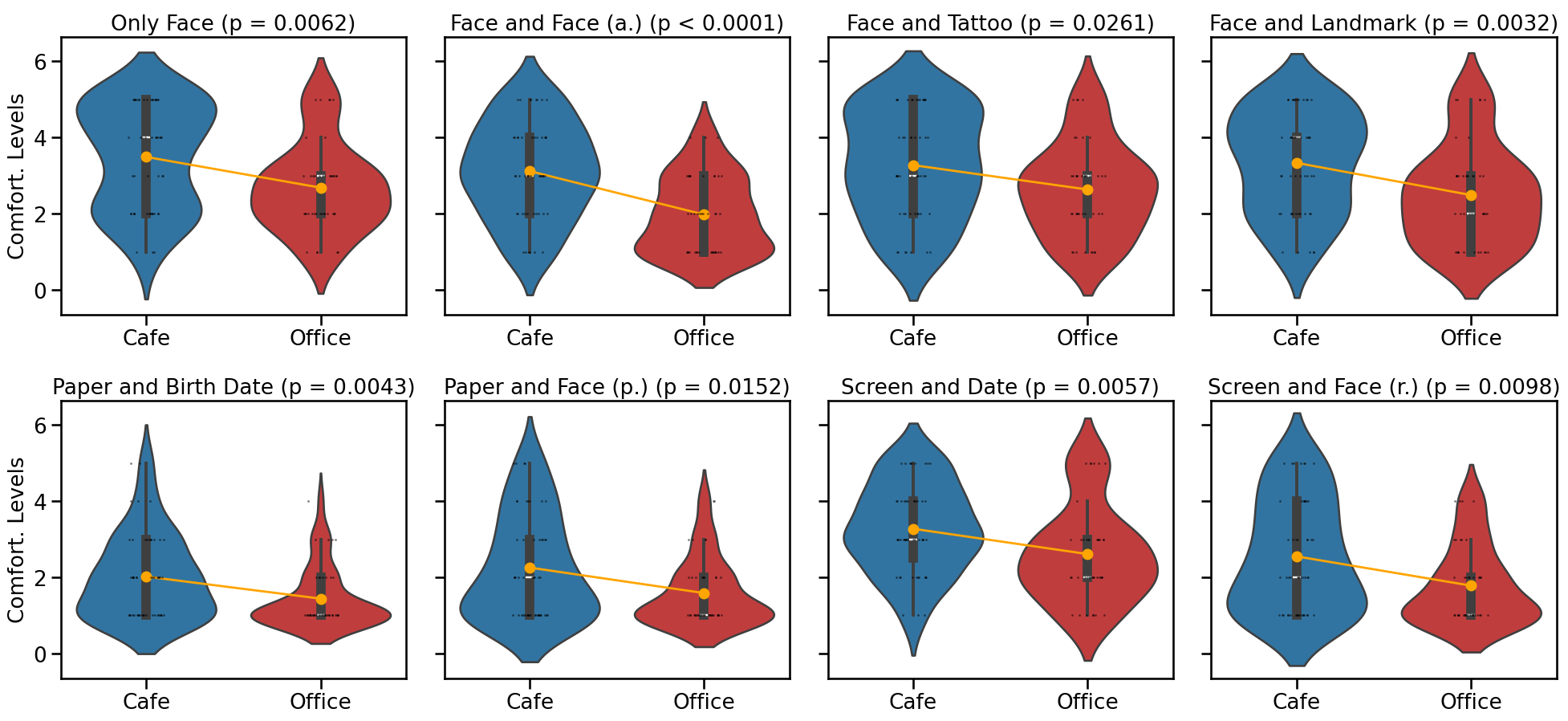}
    \caption{Violin Plots of significant Mann-Whitney U test results for café and office environments (\hyperref[rq:env]{H2}). Each subplot shows comfortability levels (1–5) for a foreground/background PSO pair; width reflects data density, and orange lines mark group mean differences.}
    \label{fig:sig_env}
\end{figure*}

\end{document}